\documentclass[12pt]{article}%
\usepackage{graphicx}
\usepackage{float}
\usepackage{subfig}
\usepackage{amsmath}
\usepackage{amsfonts}
\usepackage{multirow}
\usepackage{setspace}
\usepackage{array}
\usepackage{amssymb}
\usepackage{lscape}
\usepackage[authoryear]{natbib}
\usepackage{color}
\usepackage{color}
\usepackage[usenames,dvipsnames]{xcolor}
\usepackage{hyperref}
\usepackage{breakurl}
\usepackage{rotating}%
\usepackage{tikz}
\newtheorem{proposition}{{\bf \sc Proposition}}

\def \eproof{\hbox{\hskip3pt\vrule width4pt height8pt depth1.5pt}}

\begin{document}

\title{Diffusion in Networks and the Unexpected Virtue of Burstiness}
\author{Mohammad Akbarpour\thanks{Graduate School of Business, Stanford University,
Stanford, California 94305-6072 USA.
Email: \href{mailto:mohamwad@stanford.edu}{mohamwad@stanford.edu}.
Financial support from the Becker-Friedman Institute of the University of Chicago.}
\hspace{7mm} Matthew O. Jackson\thanks{Department of Economics (Stanford
University, Stanford, California 94305-6072 USA) and external
faculty member at the Santa Fe Institute and a fellow of CIFAR. Email:
\href{mailto:jacksonm@stanford.edu}{jacksonm@stanford.edu} .
Financial support from the
NSF under grant SES-1155302 and from grant FA9550-12-1-0411
from the AFOSR and DARPA, and ARO MURI award No. W911NF-12-1-0509.
We thank Shayan Oveis Gharan for helpful conversations and Songyuan Ding for coding some of the simulations reported here.}}
\date{Draft: December 2017}
\maketitle

\begin{abstract}

Whether an idea, information, infection, or innovation diffuses throughout a society depends not only on the structure of the network of interactions, but also on the timing of those interactions.
Recent studies have shown that diffusion can fail on a network in which people are only active in ``bursts'', active for a while and then silent for a while, but diffusion could succeed on the same
 network if people were active in a more
random Poisson manner.  Those studies generally consider models in which nodes are active
according to the same random timing process and then ask which timing is optimal.  In reality, people differ widely in their activity patterns -- some are bursty and others are not. Here we show that, if people differ in their activity patterns, bursty behavior does not always hurt the diffusion, and in fact having some (but not all) of the population be bursty significantly helps diffusion.  We prove that
maximizing diffusion requires heterogeneous activity patterns across agents, and the overall maximizing pattern of agents' activity times does not involve any Poisson behavior.



\end{abstract}

\setcounter{page}{0}\thispagestyle{empty} \newpage

\section{Introduction}

Networks of interactions are the backbone of a range of diffusion processes from the adoption of new technologies (e.g., \cite{rogers1995,banerjeecdj2013}) to the spread of ideas, behaviors, and
diseases (e.g.,  \cite{pastor-satorrasv2000,lopez2008,aralms2009,centola2010,jacksony2011,aralw2012,
akbarpour2016,araln2017,gleesond2017}).
Diffusion and contagion processes are shaped not only by the structure of the links within a network, but also by the time patterns at which links and nodes are active.
The timing of interactions in many networks are far from being time-independent.   For example, the ``burstiness'' of the timing of interactions has been documented in a multitude of diffusion processes, from email and phone conversations to gene expressions \citep{johansen2004probing,wu2010evidence,barabasi2011,holmes2012,pfitzner2013,porterg2016,onaga2017}.

In this paper we provide a theoretical analysis of how the timing of interactions affects a diffusion process.
We show that heterogeneity in activity patterns across agents actually increases the expected reach of diffusion processes.
Although previous studies have found that the timing of interactions affects diffusion process, those analyses have generally considered homogeneous populations and varied
the whole population's activity pattern \citep{belykh2004blinking, vazquez2006, vazquez2007, iribarren2009impact, pan2011path,  barabasi2011, karsai2011small, hoffmann2012generalized, scholtes2014, bick2015asynchronous, li2016fundamental, holme2012temporal, porter2014dynamical}.
However, in fact people differ widely in the timing of their active periods.     Some people check email on a very frequent and intermittent basis, while others have greater time between activity but then spend a longer time active once they are.
To date, nothing is known about how such heterogeneity influences diffusion.

We examine how combinations of time patterns of interactions affect the extent of diffusion.
The model that we examine is a variation on the widely-studied SIR model \citep{bailey1975}, which has its roots in the Reed-Frost model (see \citep{jackson2008} for background).  Some node of a network is the first infected with a disease or idea.  The infection then spreads at random through the network.
Nodes are either infected or susceptible. They begin as all being susceptible and become infected if they interact with a contagious neighbor.   Once infected, agents are contagious for $T$ periods and then cease to be contagious. Thus, diffusion spreads by having an infected and contagious node interacting with any of its neighbors who are susceptible.
What distinguishes our model is that the probability that a node is active is not independent of time.
On average, nodes are randomly active during any given period with a probability $\lambda>0$, but the probability is not independent
of the history of that node's past behavior.
Nodes' active times follow a Markov chain:  the probability that a node is active in one period depends on whether it was active last period.
In addition, nodes can differ in their Markov processes.   Some nodes are more likely to be active if they were active last period, while others are less likely to be active if they were active in the last period.   Thus, they can differ in their serial correlation patterns.
We emphasize that we still maintain that the timing of activity is independent {\sl across} nodes and nodes must all have the same
average level of activity - so that every node is active a fraction of $\lambda$ of all periods.   The key novelty in our model is allowing different nodes to have different time-dependencies in their behaviors.

Our main results illustrate that configurations of nodes that maximize the extent of diffusion as well as the probability of an epidemic are those that
have different Markov chains for different nodes.
We show that it is never maximizing to have all nodes follow the same Markov chain:  heterogeneity is {\sl necessary} to maximize diffusion.
We also fully characterize the maximizing structure of heterogeneity for a few
simple networks such as chain and star networks, providing
the basic intuition as to why it is useful to have heterogeneity and illustrating that it can help.
Combining nodes with extreme positive autocorrelation (``Sticky" nodes)  with others who have extreme negative autocorrelation (``Reversing'' nodes)
is optimal in such simple networks.
As a by-product, this
also shows which structures minimize diffusion (generally homogeneous ``Sticky'' nodes).
Depending on the application, one may wish to maximize or minimize diffusion.
Regardless of what one wishes to do, understanding how heterogeneity matters is essential for shaping policy.
As the general problem of characterizing the optimal structures for complex networks appears intractable, we analyze a couple of others by simulation.

\begin{figure}[t]
\centering
\includegraphics[width=0.5\textwidth]{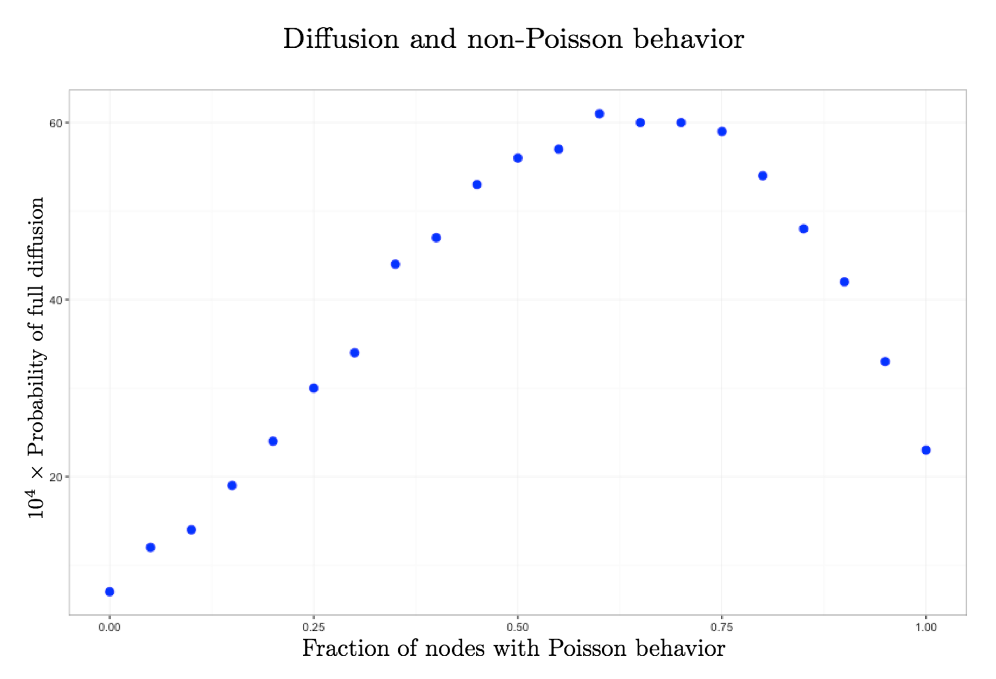}
\caption{\label{figheterog} The probability that all nodes get infected as a function of the fraction of Poisson nodes in an Erd\H{o}s-Renyi random network, when agents are either Poisson (\emph{i.i.d.} active in each period) or `Sticky' (they stay in the same state they were in the starting period, with a random starting state). When around 40\% of nodes are Sticky and the rest are Poisson, the likelihood of a full diffusion is 3 times more than when all nodes are Poisson. Clearly, mixing bursty behavior with non-bursty behavior substantially helps the diffusion. See the supplementary appendix for details behind the simulation.}
\label{fig:heterog}
\end{figure}

Figure \ref{fig:heterog} illustrates how much of a difference having heterogeneity can make.
 We examine diffusion on an Erd\H{o}s-Renyi random network (a network on $n$ nodes where there is an edge between any two nodes with probability $p$, independently across pairs.) in which agents are either Poisson (they are active each period with probability $\lambda$) or Sticky (they are either active in all periods with probability $\lambda$, or inactive in all periods with probability $1-\lambda$).  Figure \ref{fig:heterog} shows how the probability of all nodes becoming infected behaves as we vary the relative fraction of Poisson and Sticky nodes (and similar results hold for the expected
 fraction of infections).   Consistent with the previous literature, if all nodes are Sticky, diffusion is less likely than when all nodes are Poisson.
  However, when we allow agents to have heterogeneous behavior, the likelihood of full diffusion is maximized when some agents are Sticky and some are Poisson.


To understand why heterogeneity increases diffusion, consider an agent
who has recently been infected (and was just active). To maximize the chance of
diffusion, it is best if this agent behaves in a positively correlated way -- a sticky way --
so that she is more likely to remain active during the immediate
periods after infection, while she is contagious. On the other hand,
when an agent is not yet infected, it is best to
alternate states more frequently to enhance coordination probabilities.
So, sticky agents are poor receivers but good senders, and reversing agents are good receivers but poor senders.
The key is that the gain from matching a good sender and a good receiver outweighs the loss from putting together a poor sender and a poor receiver, since both matchups happen when we mix agents.
It is optimal to alternate the two types of agents.  Mixing sticky and reversing agents maximizes the sending advantage of the sticky agents and the receiving advantage of the reversing agents, without much loss from the receiving disadvantage of the sticky agent facing a reversing sender.

This is not a small effect: as seen in Figure \ref{fig:heterog}, changing just over half of the agents in a Poisson population to be sticky increases the chance full contagion by almost a factor of three in a uniformly random network.  (We see similar orders of magnitude for getting a fraction of nodes infected, and also explore other metrics for comparison in the supplementary materials.)

An analogy is helpful in getting the intuition. Imagine two people who are lost in a city with no way to communicate.  They understand that it would be best for them to find each other by trying to meet at one of the major landmarks.      To keep things simple, imagine that the city is New York and they each expect that the logical meeting places are either the Empire State Building or the Statue of Liberty, as in the seminal discussion of focal points by \cite{schelling1960}.     If they both go to each of the landmarks, then there is a chance that they will miscoordinate (\cite{chassang2010fear,kempest2016}) - going in the opposite order and thus missing each other. If instead, one of them just goes to one of the landmarks and stays there, while the other alternates and goes to both, then they are sure to meet.
Of course, they need to coordinate on who follows which strategy - that is, who stays put and who searches.   Nonetheless, the point is that a population in which people have a diversity of interaction patterns can lead to superior coordination probabilities.
While this example is extreme, it illustrates how having the two individuals use different actions can improve the probability that they interact. We show that this intuition extends to the network setting.

Our results also have some broader implications.
 Diffusion processes are critical in many economic settings from the diffusion of a new technology, to learning about a new program, to the spread of a crippling computer virus.
 For instance, in a decentralized market it suggests that having `sticky' nodes, serving as dealers, could enhance the efficiency of the movement of assets. 
 As another example, in a market where agents search for information (e.g., \cite{duffie2009information}),  existence of agents whose search patterns are `sticky' could enhance the efficiency of information percolation, and incentivizing some agents to be stickier could enhance welfare.
To further illustrate some of those we discuss an online advertising setting in which users choose between different competing websites (e.g., news agencies). If an advertiser must choose which website to buy ads on at various times, with the goal of maximizing the probability of reaching a user, then our results suggest that the advertiser should identify how users behave and then behave in the opposite way: If users are Sticky (i.e. they visit a specific news website and stick to it), then the advertiser should alternate between websites to reach the users.  However, if users alternate between websites, then the advertiser should stick to one website so as to minimize the probability of miscoordination.

\section{The Model}

There are $n\geq 3$ agents, with labels $i\in N=\{1,\ldots,n\}$ connected in a network represented by a simple graph $G = (N,g)$, where $g \subseteq N^2$ and $ij \in g$ if agent $i$ and agent $j$ are \emph{linked}.

Time passes in discrete periods $t\in \{1,2, \ldots\}$.  (We work in discrete time to easily admit negative autocorrelation.  For this problem, discrete time
seems to simplify rather than complicate the analysis and intuitions.)
Agents are either \emph{active} or \emph{inactive} in a given period. Activity is independent across agents.
An agent is active with a probability $\lambda\in (0,1)$ in any given period, on average.
We assume the long-run average activity levels are the same for all agents.  By focusing on agents who are homogeneous in
\emph{how often} they participate,  we can focus on
the effect of heterogeneities in the \emph{timing} of participation on diffusion.

\begin{figure}
\centering
\begin{tikzpicture}[inner sep=0]
\tikzstyle{every node}=[circle,draw, minimum size=15mm];
\node at (-2,0) (b) {$inactive$};
\node at (2,0) (c) {$active$};
\draw[->,line width=1pt]
  (b)  edge [bend left=40] (c)
  (c) [bend left=40] edge (b);
\tikzstyle{every node}=[circle];
\node at (0, 1.5) {$q_i$};
\node at (0, -1.5) {$p_i$};
\node at (2, 2.3) {$1 - p_i$};
\node at (-2, 2.3) {$1 - q_i$};
\path[line width=1pt]
(c) edge [loop above]  (c)
(b) edge [loop above] (b);
\end{tikzpicture}
\caption{Activity Markov Chain of agents}
\label{fig:MarkovChain}
\end{figure}
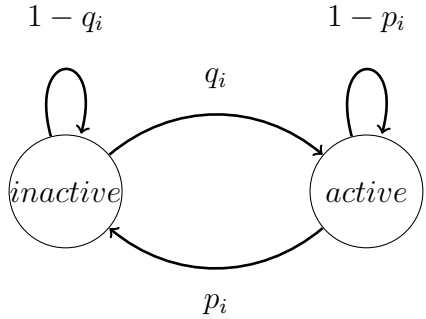

In particular, an agent's activity follows a Markov chain.  If an agent $i$ is active in period $t$, then s/he is inactive in period $t+1$ with probability $p_i$,  and active with probability $1-p_i$.
Similarly, if an agent $i$ is inactive in period $t$, then s/he is active in period $t+1$ with probability $q_i$,  and inactive with probability $1-q_i$.  This is pictured in Figure \ref{fig:MarkovChain}.

For any fixed $\lambda$, an agent is then completely characterized by $p_i$, or equivalently by $q_i$.
In particular, the following equality must hold:
\[
\lambda p_i = (1-\lambda)q_i,
\]
which is just the usual balance equation of the Markov chain, given that $\lambda$ is the steady-state probability of activity.
Some useful rewritings of the this equation are:
\[
\frac{\lambda}{1-\lambda} = \frac{q_i}{p_i}
, \ \ \ \ \ \
\lambda = \frac{1}{1+\frac{p_i}{q_i}}
, \ \ \ \ \ \
q_i=p_i \frac{\lambda}{1-\lambda}.
\]

So, our agents are completely described by $p_i$, given any fixed $\lambda$.

\subsection{Three Benchmark Types}

\
There are three levels of autocorrelation that serve as benchmarks.

A \emph{Poisson} agent is one who has $p_i=1-\lambda=1-q_i$.  This is an agent who is active at every period with probability $\lambda$; that is, her state is i.i.d. over time.

A \emph{Sticky} agent is one who has $p_i$ and $q_i$ both `near' 0.   This is an agent whose state is (almost) perfectly autocorrelated over time.
In particular, let \emph{Sticky} agents be those who are either always on (with probability $\lambda$), or always off (with probability $1-\lambda$). So this is the limit of a Markovian agent as $\min[p_i, p_i \frac{\lambda}{1-\lambda}]\rightarrow 0$, but one that is degenerate.

A \emph{Reversing} agent is one with the maximal possible $p$ and $q$ (maximal negative autocorrelation):  so $p=1$ if $\lambda \leq 1/2$ and
$p= (1-\lambda)/\lambda$  if $\lambda \geq 1/2$.    Similarly,  $q=1$ if $\lambda \geq 1/2$ and
$q=\lambda/ (1-\lambda)$  if $\lambda \leq 1/2$.
Thus, the state of a reversing agent is as negatively serially correlated as possible, switching back and forth between being active and inactive as frequently as possible.
In the case in which $\lambda=1/2$, a Reversing agent simply reverses its state every period.

To understand these three types, let us consider workers in a large firm who spend $\lambda$ of their time near their office where their door is open for interaction with other co-workers on their team, and then $1-\lambda$ of their time working on projects with closed doors or away from their office.  Let us think of periods as hours. Sticky agents are those who schedule their closed door time in large clumps, so that they are unavailable for some number of hours in a row, then available with an open door for some number of hours in a row, etc., so that they schedule their sequestered work time in contiguous segments spending days in a row on a project, then days in a row available in the office.  If they are available with an open door at some point, they are more likely to be available the next hour, and conversely if they are occupied or away then that is also likely to persist.  
Poisson agents would be people who just randomly schedule closed door project time with no particular pattern.  
Reversing agents would be people who prefer to alternate, so they work on a project for an hour, then open their door and interact for an hour, then close their door and work for an hour, etc.; alternating project time with interaction time.  

We can think of Poisson, Sticky, and Reversing agents as the canonical cases:  one with no autocorrelation, one with maximal positive autocorrelation, and the other with maximal negative autocorrelation.
Of course, there are other levels of autocorrelation in an agent's state, and we admit arbitrary cases in our general analysis.

\subsection{Diffusion}

Some agent is initially infected.  All other agents are initially susceptible.  Once an agent becomes infected, the agent stays infected forever after.
An agent can transmit infection for $T$ periods after being infected.  We say that such an agent is `contagious' during those time periods.
In each period, an agent who is contagious transmits the infection to a neighbor if and only if both he and his neighbor are active, and his neighbor is susceptible.

\section{Line Networks and Canonical Agents}

\begin{figure}
    \centering
    \begin{tikzpicture}
  [scale=0.8, auto=left,every node/.style={circle, minimum width=0.7cm, fill=blue!20}]
  \node (n6) at (0,0) {};
  \node (n4) at (2,0)  {};
  \node (n5) at (4,0)  {};
  \node (n1) at (6,0) {};
  \node (n2) at (8,0)  {};

  \foreach \from/\to in {n6/n4,n4/n5,n5/n1,n1/n2}
    \draw (\from) -- (\to);
\end{tikzpicture}
    \caption{A Line Network with Five Nodes}
    \label{fig:Line}
\end{figure}
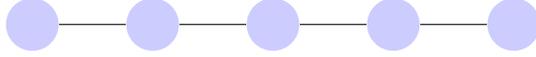

We begin our analysis by looking at networks that are `lines'  - a tree in which no agent has degree more than two. Figure \ref{fig:Line} is a line network with five nodes.
These networks illustrate the main ideas and intuitions and permit a complete characterization of the maximizing configurations when we restrict our attention to the canonical agents.

We begin with an analysis of diffusion with only Poisson and Sticky agents - as these are sufficient to provide the basic intuitions about how heterogeneity helps with improving diffusion.   After establishing results on
optimal configurations with these types, we then add in the Reversing agents, showing that optimal configurations mix the extreme agents:  Sticky and Reversing agents.
Finally, we turn to an analysis with general agent types.

\subsection{Poisson and Sticky Agents}

Let $PS$ denote the probability that a Poisson agent who is infected transmits to a sticky neighbor who is susceptible within $T$ periods; and similarly define $SP$, $PP$, $SS$.
Similarly, let $PPP$ denote the probability that there is full transmission among three Poissons in a line - where transmission must occur within a new $T$ periods for each successive interaction.   So, once infected, a person can pass the disease or idea along to a neighbor for $T$ periods from the date of the current agent's first infection.   Likewise, we define $PSP$, and so forth.  We abuse notation and occasionally also use this notation to refer to a particular configuration of agents.

When we consider the transmission in a line we presume that the initially infected agent is a node at one end and that the agent is then randomly active with probability $\lambda$ in the first period.  If we instead assumed that the first agent begins by being active, then the first agent should always be Sticky, and that would just push the whole problem back one agent.

In all of the analysis that follows, we presume that agents have the same overall probability of being active, but differ only in timing.   Our interest is in seeing how the patterns of timing matter, and holding constant the overall level of activity allows us to isolate how patterns of autocorrelation matter.

\begin{proposition}
\label{three}
Consider 3 agents in a line, with all agents being independently active with probability $\lambda\in (0,1)$ in steady state, and who once infected
can transmit for some positive integer number of periods $T$.
The configuration of Poisson and Sticky agents that maximizes both the expected number of infections and the probability that all agents become infected is uniquely:
\begin{itemize}
\item $PSP$
if $\lambda< \lambda^*$,  and
\item $PPP$
if $\lambda > \lambda^*$,
\end{itemize}
where $\lambda^*$ is the unique solution in $(0,1)$ to:
\begin{equation}
\label{psp}
\lambda =  \left[\frac{ 1- (1-\lambda^2)^T}{ 1- (1-\lambda)^T} \right]^2.
\end{equation}
if we are maximizing the probability of total infection, and
\begin{equation}
\label{psp2}
\lambda=  \frac{ \left[ 1-(1-\lambda^2)^T\right]\left[ 2-(1-\lambda^2)^T\right]}{ \left[ 1-(1-\lambda)^T\right]\left[ 2-(1-\lambda)^T\right]} .
\end{equation}
if we are maximizing the expected number of infected nodes. 
(For any $T$, there is a unique fixed solution to (\ref{psp}) and (\ref{psp2}) in $(0,1)$, as we show in the proof.  1 and 0 are also solutions, but uninteresting ones, as then agents are either always or never active, in which case the time series of their activity is irrelevant.)
The interior solution of (\ref{psp2}) is smaller than that of (\ref{psp}).
\end{proposition}

The proof of this proposition, as well as the proof for all other propositions and theorems are in the appendix.

\medskip
To understand the trade-offs that drive heterogeneity note that once an agent is infected, it is best to have that agent be Sticky because a recently infected Sticky agent remains active while she is contagious. However, when an agent is not yet infected, it is best to alternate states randomly, to enhance coordination probabilities.   Sticky agents are poor receivers but good senders, and Poisson agents are good receivers but poor senders. Such dynamics make it optimal to connect a Sticky sender and a Poisson receiver.
Moreover, the probabilities of transmission have synergies - matching Sticky senders with Poisson receivers increases overall probability more than the subsequent loss due to then having to subsequently alternate a Poisson sender with a Sticky receiver.
Under a wide range of activity levels, the advantages of having heterogeneity outweigh the losses from having the receiver be Sticky.

The point of considering the ``optimal'' configuration is not necessarily to suggest that there is some mechanism designer or planner
who can control the system, but to show that heterogeneity enhances diffusion in certain contexts and to understand why this occurs -- by showing that it maximizes diffusion, we can see that it definitely enhances diffusion.
Our simulations will also show that this is not a small effect.

We now show that this intuition extends to longer lines.

\begin{proposition}
\label{tree}
Consider an odd number of agents in a line,  with all agents being independently active with probability $\lambda\in (0,1)$ in steady state, and who once infected
can transmit for some positive integer number of periods $T$.
Start with one end node being infected and let   $\lambda^*\in (0,1) $ solve (\ref{psp}) and
$\lambda^{**}$ be the interior solution of  $\lambda=\left[ 1- (1-\lambda)^T\right]^2$.
Then $0<\lambda^{**}<\lambda^*<1$ and
the configuration that maximizes the probability of overall infection is:
\begin{itemize}
\item $PSSSS\ldots SP$ if  $\lambda<\lambda^{**}$,
\item $PSPSP\ldots SP$ if $\lambda^{**}<\lambda< \lambda^*$, 
and
\item $PPPPP\ldots PP$ if $\lambda>\lambda^*$,
\end{itemize}
\end{proposition}

As we saw in Proposition \ref{three}, there are similar results for the case of maximizing the expected extent of the diffusion.
The cutoff expressions become more complex with longer lines, and so in Proposition \ref{tree} we simply provide the analysis for the probability of overall infection.  We can still see the gain from heterogeneity in the following simulation.

To see the extent of the gain from alternation, consider the following results from simulations.
We compare the infections in a line of five nodes in which all of the nodes are Poisson to one in which they alternate Poisson and Sticky.   One of the nodes is picked at random to be infected and we set  $T=2$.
We show the comparisons for a full range of $\lambda$.
For each of 50 values of $\lambda$ we run 40000 iterations of drawing a random network and running an infection.  The reported values for each $\lambda$ value are the average over the 40000 iterations.

\begin{figure}[t]
\centering
\subfloat[]{
\label{fig:ChainsRatio5}
\includegraphics[width=0.45\textwidth]{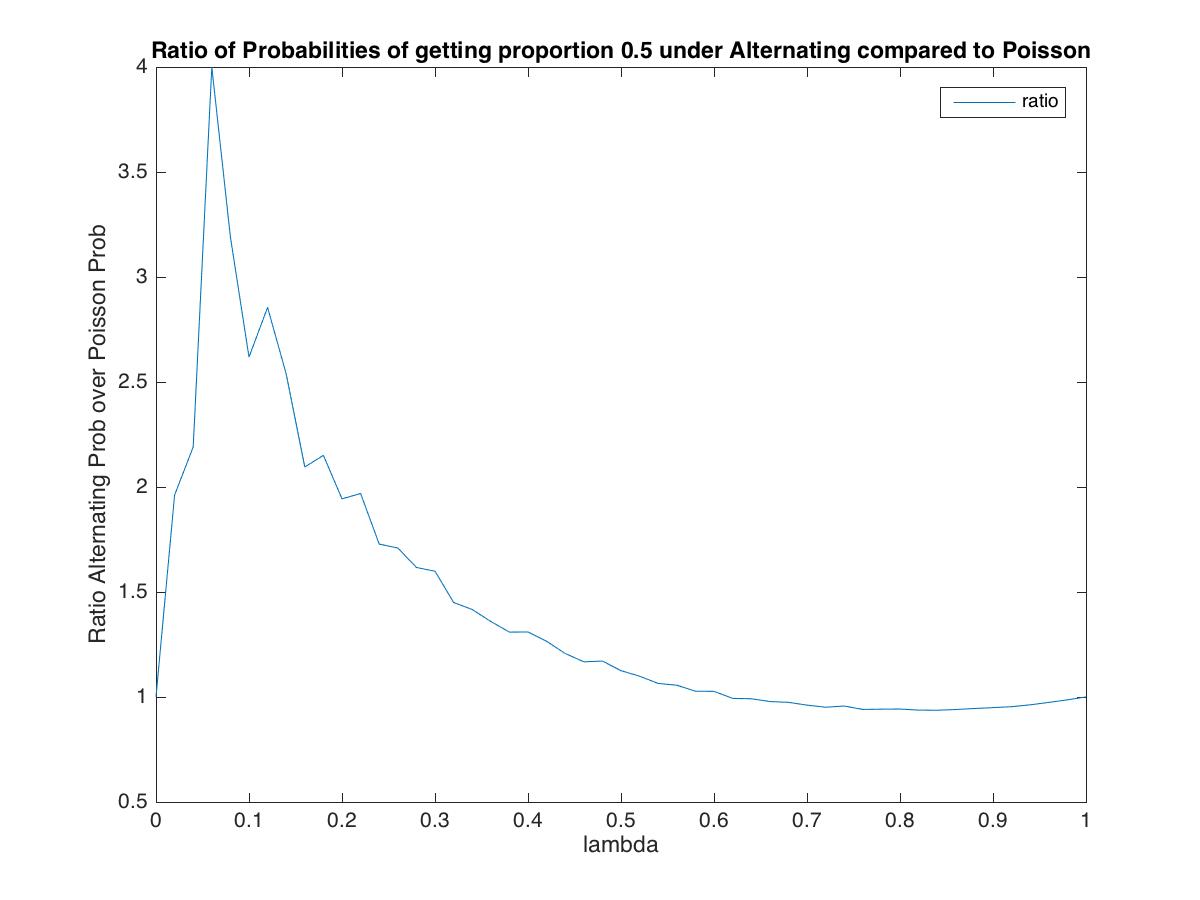}
}
\subfloat[]{
\label{fig:ChainsRatio1}
\includegraphics[width=0.45\textwidth]{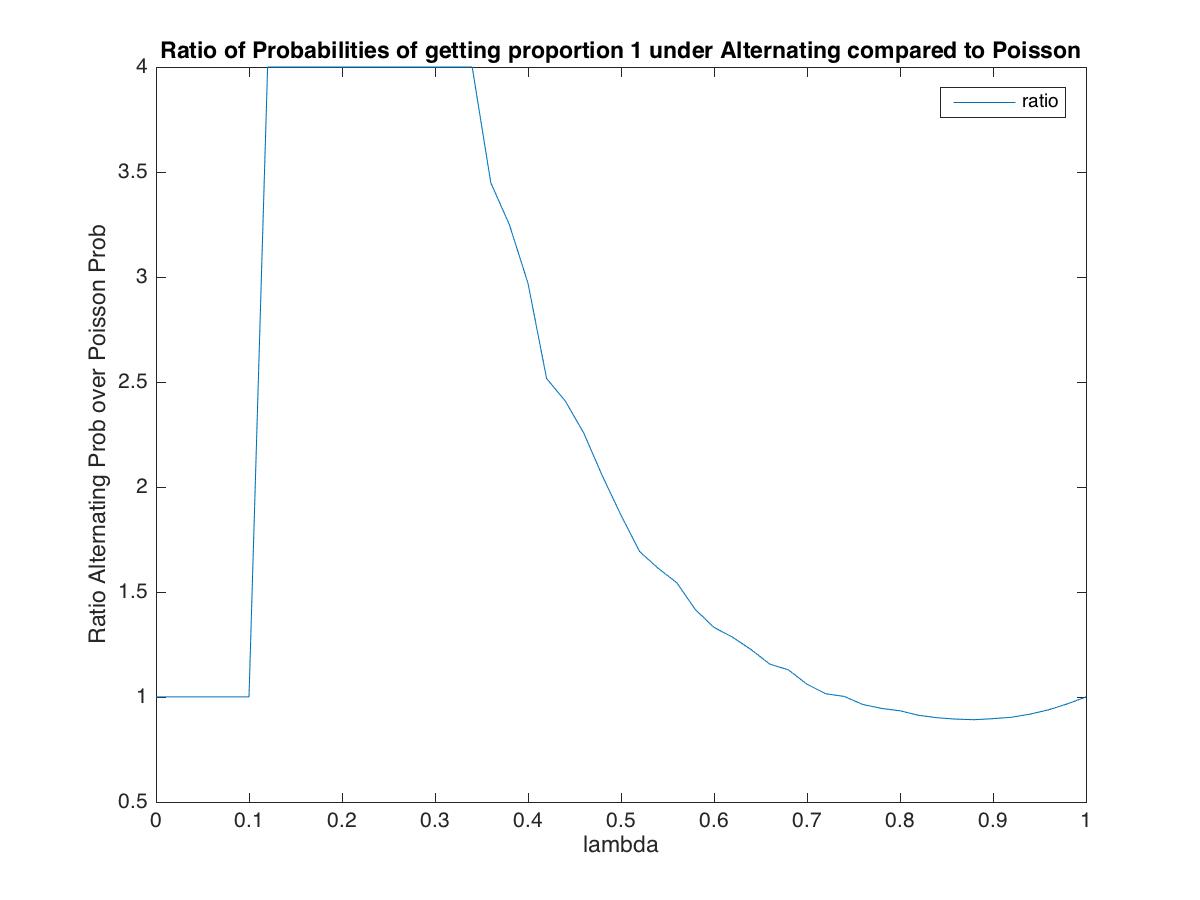}
}
\caption{\label{figChains3} The ratio of the infection probabilities under alternating Poisson and Sticky nodes over that for all Poisson Nodes, on chains of five nodes with one randomly infected: (a) ratio of probability of getting at least half of nodes infected, and (b) Ratio of probability of getting all nodes infected.
}
\end{figure}

We see in Figure \ref{figChains3} that the gains from alternating Sticky with Poisson compared to having just Poisson can be very large, more than four hundred percent,  while the reverse advantage that comes in at high levels of $\lambda$ is relatively negligible.

In the introduction, we discussed how recent studies had showed that bursty behavior slows down the diffusion (see, for instance, \citep{vazquez2006}). You may ask whether our results contradict those findings, and the answer is no. In our setting, $SS \cdots S$ is dominated by $PP \cdots P$, which confirms those results. What we prove, however, is that one should not conclude that all Poisson is optimal, as
 those results did not admit heterogeneity.  When we consider all possibilities (as we also saw in Figure \ref{fig:heterog}), mixing sticky/bursty behavior with non-bursty behavior substantially improves the diffusion.

We next show that maximizing the probability of a full infection tilts the balance more towards Sticky nodes at key junctures or ``hubs''. To expand on this point, we study the diffusion process on a ``star" network. A star network has a central node and $n$ leaves connected to the center. For example, in Figure \ref{fig:Star} we see a star network with six leaves.

 \newcommand*\ab{.35}
  \tikzset{
    net node/.style = {circle, minimum width=2*\ab cm, inner sep=0pt, outer sep=0pt, fill=blue!20},
    net root node/.style = {net node, minimum width=2*\ab cm},
    net connect/.style = {line width=1pt, draw=black},
  }
  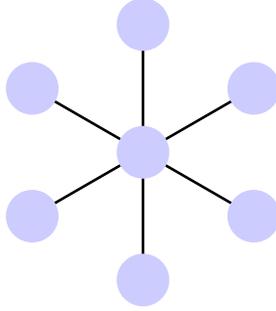
\begin{figure}
    \centering
    \begin{tikzpicture}
      \node (root) [net root node] {};
      \foreach \i in {0,...,5}
        \path [net connect] (root) -- (-90+\i*60:1.7) node [net node] {};
    \end{tikzpicture}
    \caption{A star network with six leaves}
\label{fig:Star}
  \end{figure}

\begin{proposition}\label{star}
Consider agents in a star network with $n$ leaves, with all agents being independently active with probability $\lambda\in (0,1)$ in steady state, and who once infected
can transmit for $T = 2$ periods.  Start with some random leaf being infected. Then, for  any $\lambda$, there exists some $N$ such that if $n \geq N$, then the configuration that maximizes the probability of a full contagion is to have an $S$ node in the center and $P$ nodes on the leaves.
\end{proposition}

To see why Proposition \ref{star} is true,
note the following.  First consider whether the infected leaf infects the center node.  This does not depend on how many other leaf nodes there are.  Second, consider how the center node interacts with other nodes once infected.  Here there is an advantage to having the center be sticky, so that it can transmit in both periods once it is infected.  This advantage grows with the number of leaves.  Thus the advantage to having it be sticky as a sender grows with its degree, while the disadvantage of having it be sticky as a receiver does not depend on degree.  As degree grows, the advantage can become overwhelming as the expected gain in the diffusion from having the center be sticky compared to Poisson grows in the number of nodes. 

Proposition \ref{tree} shows that for $\lambda > \lambda^*$, the configuration that maximizes the expected number of infected nodes does not include $S$ agents.
In contrast, Proposition \ref{star} claims that including Sticky behavior is \emph{always} optimal for agents at sufficiently central junctures. Note that ``hubs'' appear in various kinds of networks, from human brain [\citep{bullmore2009complex, van2013network}] to social networks [\citep{kempe2003maximizing}] to computer networks [\citep{cohen2003efficient}], and this proposition suggests that in designing the activity patterns of such nodes with high degrees, autocorrelated behavior can be optimal.

In the next section, we show how ``extreme heterogeneity" (i.e., mixing Sticky agents with Reversing agents, as opposed to mixing Sticky and Poisson agents) further improves diffusion.

\subsection{Reversing Agents}

Continuing our comparisons, we now consider what happens when we also consider Reversing nodes.

Reversing nodes do not do so well when matched with each other, as they only happen to coordinate if they are in similar states in the first period (either active or inactive), but they can badly miscoordinate if they are in different starting states when $\lambda$ is low.   Reversing nodes, however, work very well when matched with Sticky nodes.

\begin{proposition}
\label{treeReverse}
Consider agents in line and begin with one end node infected, but then randomly active in the first period of its transmission.
Suppose that all agents are independently active with probability $\lambda\in (0,1)$ in steady state, and once infected
can transmit for some positive integer number of periods $T$.
If $\lambda<\lambda^{*}$,  then any configuration that maximizes the expected number of infected nodes or the probability of overall infection
involves $R$ nodes.

Moreover, in the case of $T=2$, then the optimal configurations involve only $R$ and $S$ nodes (Poisson nodes are not used in the optimal configurations).
Those optimal configurations are either to have full alternation of the form $RSRSR...SR$ for low values of $\lambda$,   all Reversing nodes $RRRRR...RR$ for high levels of $\lambda$,
and some combinations of string of $RRR$'s and alternating $SRSR...SR$ for middle values of $\lambda$.
\end{proposition}

Having a Reversing node following a Sticky node maximizes the probability of transmission.  For example, if $\lambda>1/2$ then the probability of transmission from an infected Sticky to a Reversing node is one.  Thus, the only loss in having alternaging Sticky and Reversing nodes is from having Sticky nodes as receivers which is biggest for large values of $\lambda$, at which point it is bettr go entirely to Reversing nodes.

\subsection{Illustrations}

Before moving to networks with cycles, we further illustrate our results on a line of five nodes.  This shows the differences between various combinations of node types and shows how much
improvement comes from including extreme node types and from heterogeneity.


We compare the infections in a line of five nodes for the following cases: all nodes are Poisson, nodes alternate Poisson and Sticky, all nodes are Reversing, and nodes alternate Reversing and Sticky . One of the nodes is picked at random to be infected and $T=2$.
We show the comparisons for a full range of $\lambda$.
For each of 50 values of $\lambda$ we run 40000 iterations of drawing a random network and running an infection.  The reported values for each $\lambda$ value are the average over the 40000 iterations.

\begin{figure}[h!]
\centering
\subfloat[]{
\label{fig:Chains25}
\includegraphics[width=0.45\textwidth]{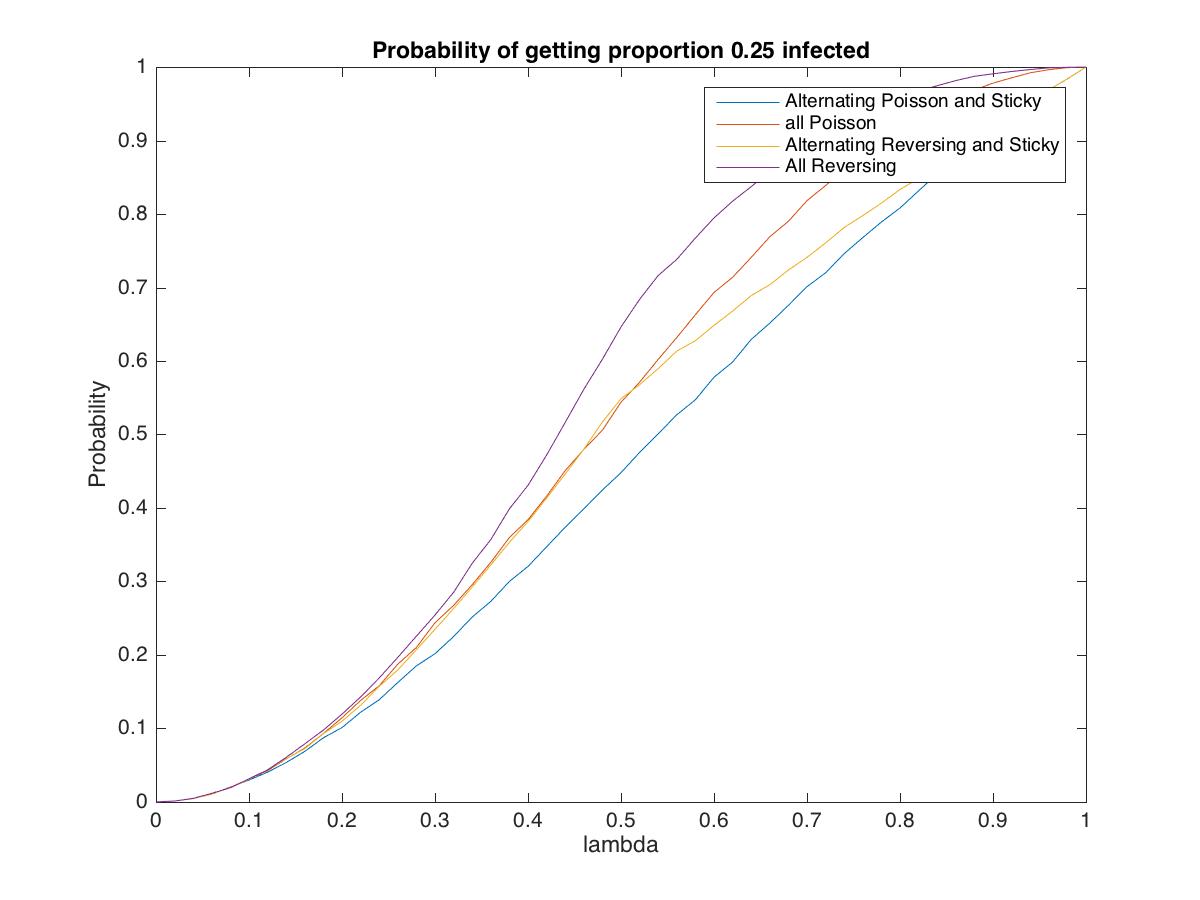}
}
\subfloat[]{
\label{fig:Chains5}
\includegraphics[width=0.45\textwidth]{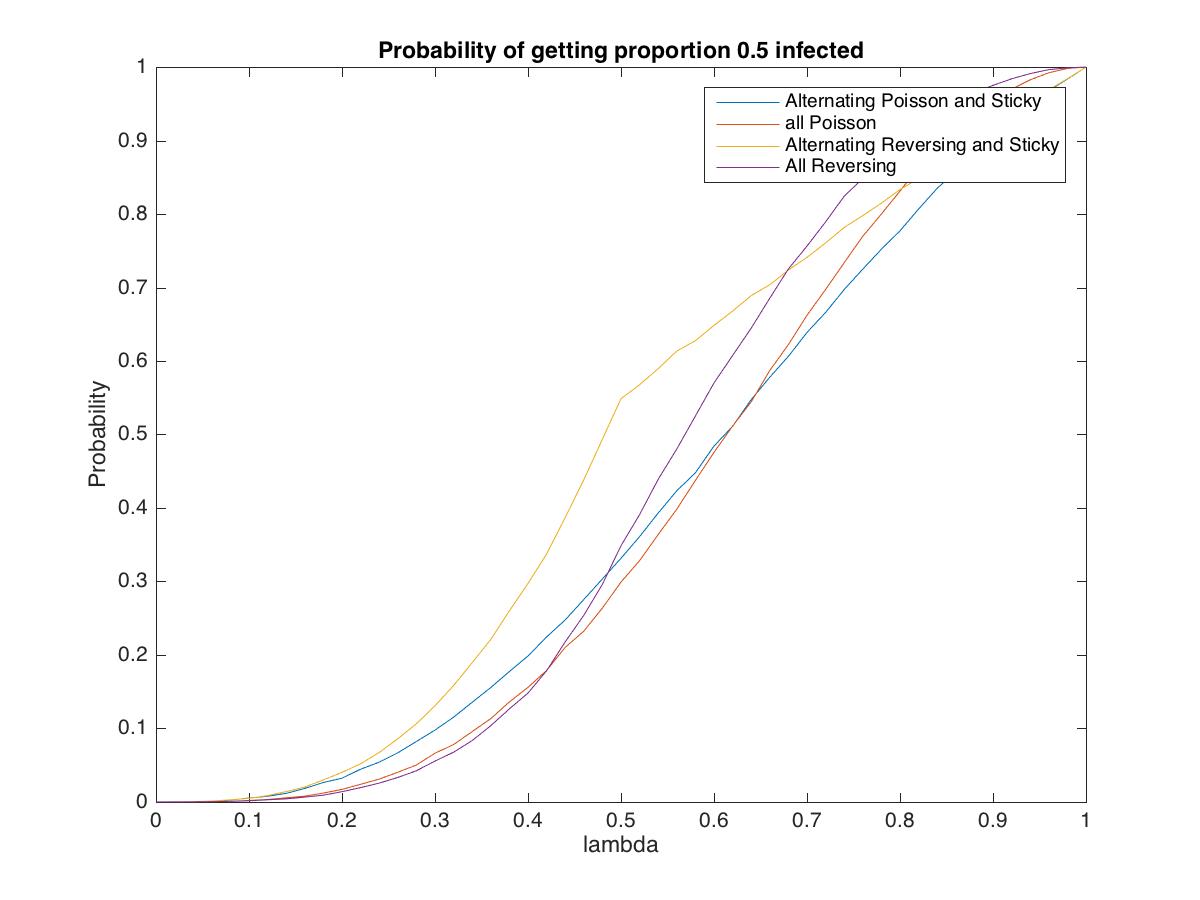}
}
\caption{\label{figChains1a}
Comparisons of infection probabilities under various configurations of nodes on lines of five nodes with one randomly infected: (a) probability of getting 1/4 nodes infected, and (b) probability of getting 1/2 nodes infected.}
\end{figure}

Figure  
\ref*{figChains1a}  shows that the best system is always either alternating Reversing and Sticky or else all Reversing - and not to involve Poisson nodes.  Again, the gains can be large in magnitude - especially for low to middle ranges of $\lambda$, in which all Poisson has a probability of about .3 of reaching half infection while alternating Reversing and Stick nodes has a probability of more than .6.

\section{General Networks}

We now move on to more general networks.  Obtaining a full characterization once one introduces cycles into a network appears intractable.
The key complication is that a node could be getting contact from several other nodes at once, and also this could be happening asynchronously.   The full array of possibilities of which nodes become
infected when and how that depends on the full network structure and activity patterns of all nodes explodes exponentially,
and renders the problem intractable.
Nonetheless, it is important to make sure that heterogeneity still matters in such networks.
So, our main goal in this section is to show that heterogeneity still plays a substantial role in more complex networks, both through a partial analytic result showing that some heterogeneity enhances diffusion in
any network, and then show via some
simulations on some richer networks, that heterogeneity makes a significant impact more generally.

\subsection{General Agents and Expected Infection Levels}

We now allow agents to have any $p_i$'s, but still where all agents have the same  long-run probability $\lambda$ of being active in order to focus on the timing patterns rather than overall levels of activity.

The following result shows that in any network that has some agents who are not in cycles the optimal configuration of agents \emph{must} involve some sort of alternation/heterogeneity.
In order to make the point that heterogeneity is always optimal, it is sufficient to consider $T=2$, as the calculations are tractable for that case.

\begin{proposition}
\label{mustalternate}
Consider any path-connected network for which there is at least one node that has degree one (a `leaf').
Suppose that each agent must be active $\lambda$ of the time, independently across agents, and
consider $T= 2$.   Start with some non-leaf node being infected.
Any configuration of $p_i$'s that maximizes {either} the expected number of infected nodes or the overall probability of full contagion involves $p_i\neq p_j$ for some $i$ and $j$.
\end{proposition}

\medskip

The proof takes advantage of some node that has degree one and its neighbor, which allows us to obtain closed form expressions for their contagion, fixing the rest of the network.  Once nodes enter into cycles, it becomes intractable to calculate the optimal configurations for nodes embedded in cycles.

As this proposition relies on leaf nodes, it becomes important to also check by simulation that heterogeneity makes a difference in more general networks.
To that end we now turn to some simulations to show that
 the basic intuition that heterogeneity in types leads to higher rates of contagion extends throughout
networks in general - and not just occasional leaf nodes - as we verify in some simulated networks.

\subsection{Random Networks}

We now examine Erd\H{o}s-Renyi networks on 20 nodes with a probability of 1/4 per link.   So, the expected degree is roughly 5 and the network is usually connected and has many cycles.
Again, we compare what happens with all Poisson nodes to what happens with half Poisson and half Sticky nodes (ten of each type), as well as having all Reversing nodes, and having half Reversing and half Sticky.  Given that the network is random, the various nodes end up randomly located in the network.
Again, for each of 50 values of $\lambda$ we run 40000 iterations of drawing a random network and running an infection.  The reported values for each $\lambda$ value are the average over the 40000 iterations.

\begin{figure}[h!]
\centering
\subfloat[]{
\label{fig:ER-Reverse25}
\includegraphics[width=0.45\textwidth]{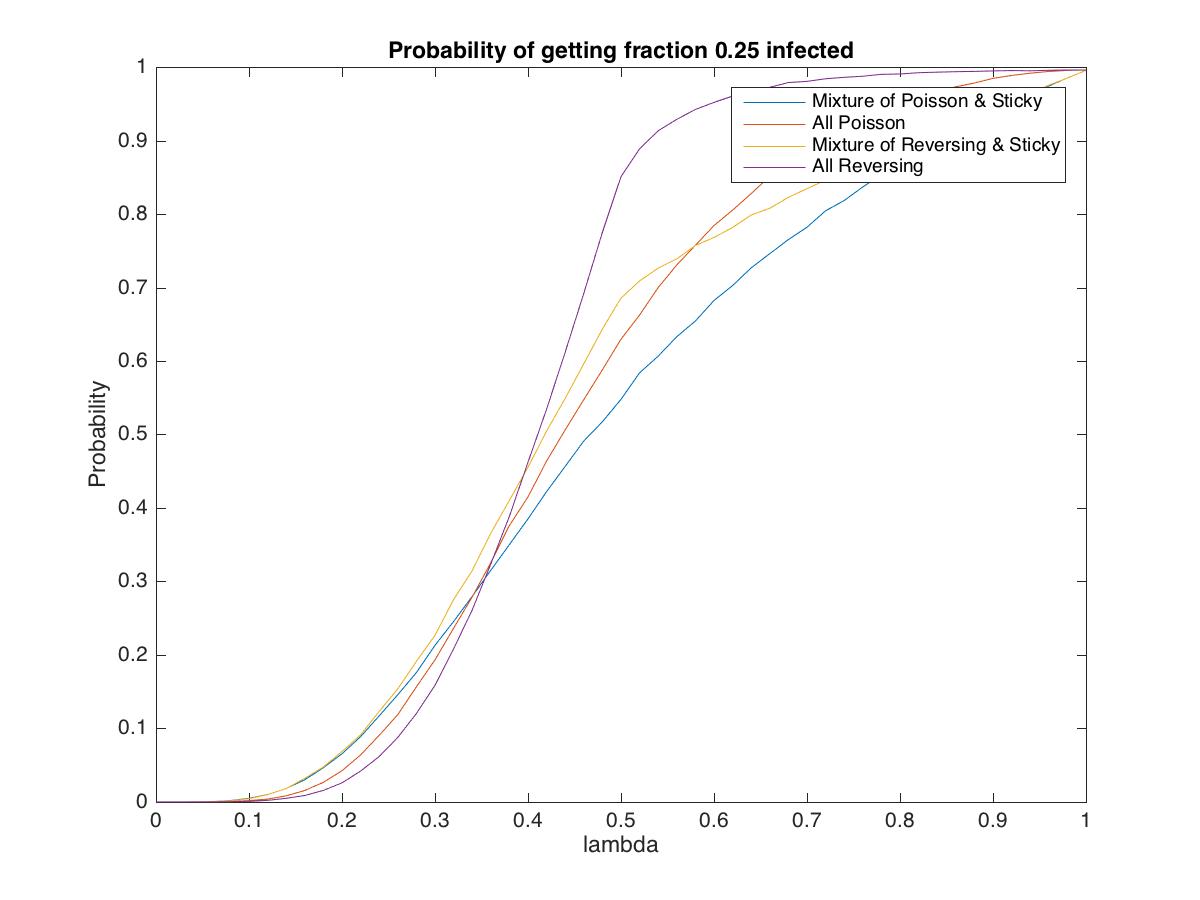}
}
\subfloat[]{
\label{fig:ER-Reverse5}
\includegraphics[width=0.45\textwidth]{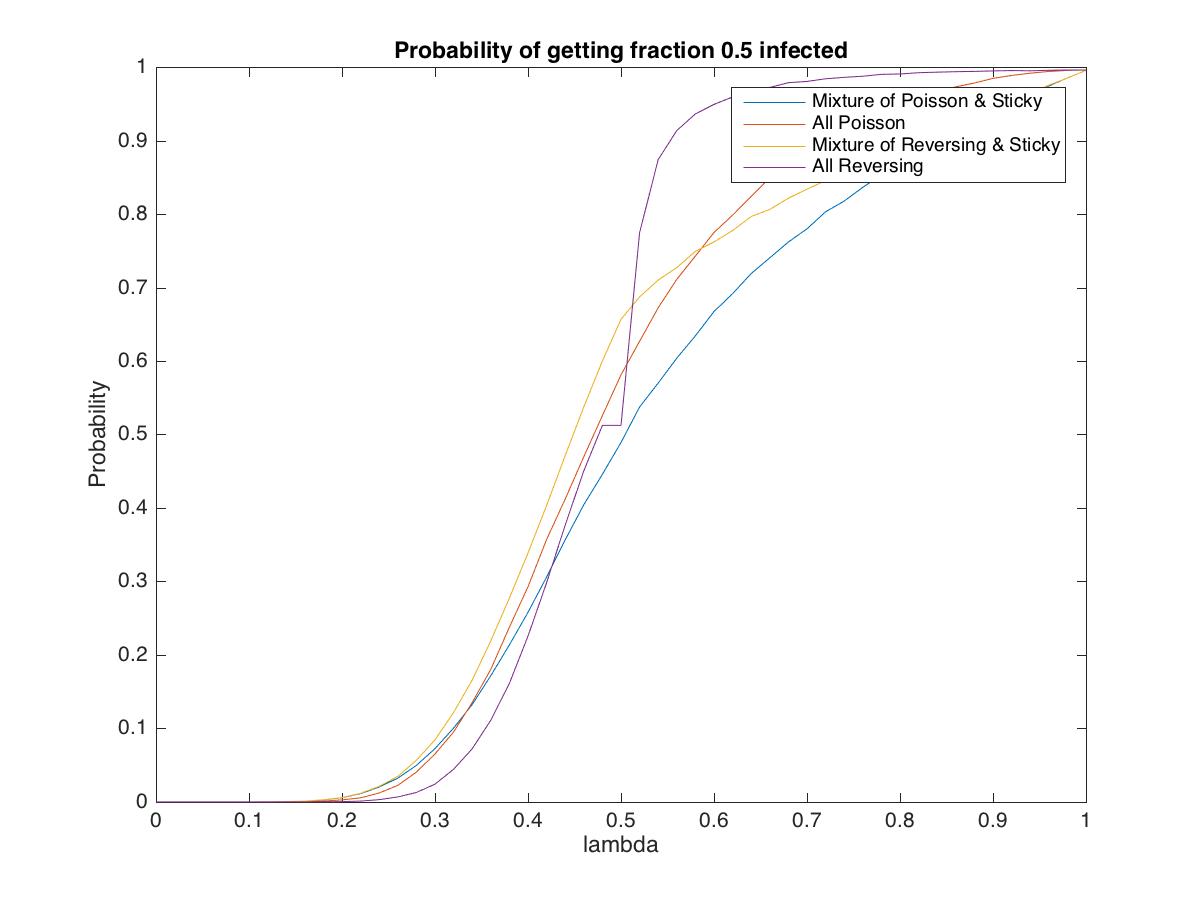}
}
\caption{\label{fig:ER-Reverse1a} Comparing all Poisson to an alternation of Poisson with Sticky to an alternation of Reversing nodes with Sticky nodes for
Erd\H{o}s-Renyi Random networks on 20 nodes: (a) probability of getting 1/4 nodes infected, and (b) probability of getting 1/2 nodes infected
}
\end{figure}

Figure  
\ref{fig:ER-Reverse1a}
 shows that a mixture of Reversing nodes with Sticky nodes does as well as either of the other configurations or substantially better for a wide range of $\lambda$ -- even for Erd\H{o}s-Renyi random networks, not just lines.
Moreover, here the nodes are not explicitly placed in some alternating fashion, but just randomly mixed in the population and still having Sticky mixed with Reversing nodes does better for a substantial range of parameters.   Having all Poisson is never optimal.

%

\section{An Illustrative Application: Online Advertising}

We point out that our results on the necessity of heterogeneity provide broader insights and, for instance, can help illuminate the design of optimal advertising strategies.

Consider a news consumer that an advertiser wants to reach, and two news websites, say BBC and CNN, that the consumer may frequent. Suppose the user is either a Sticky user, who always goes to the same news website, or a Poisson or Reversing user, who either randomly or regularly switches between the two news website.

Consider an advertiser with a fixed budget, enough for $N$ display ads. Suppose the advertiser can only put an ad on one of the two news agencies in any given day. The goal of the advertiser is to maximize the probability of reaching the consumer.  Should the advertiser alternate between BBC and CNN on different days, or she should pick one of the two and spend all of the budget at that one, or follow some other strategy?

Our results tell us that the optimal strategy for the advertiser is to follow a complementary timing to the consumer: If the consumer is Sticky, then the optimal decision is alternate (`Reverse') between BBC and CNN  so as to ensure reaching the consumer no matter which news agency she regularly visits. If the user alternates in some manner - being Poisson or Reversing, then the optimal decision of the advertiser is to pick just one of the news outlets, BBC or CNN, and spend all of the budget there, (so the advertiser behaves as a `Sticky' agent) which minimizes miscoordination chances.

\section{Concluding Remarks}

Heterogeneity in activity patterns among a population enhances diffusion, and matching extreme types of agents next to each other can increase the likelihood of diffusion substantially.
 The conclusion that bursty/sticky agents hurt diffusion only holds when one requires all agents to have the same activity patterns, and the relationship between diffusion probability and fraction of agents with bursty behavior has an inverse-U shape.
We show this analytically in simple networks.
Given the intractability of finding fully optimal configurations in general networks, we illustrated that the results hold in some more complex networks by simulation.

The point here is not to fully characterize the optimal patterns for diffusion, as it is both intractable and not clear what one would do with such a characterization.  The point instead is to show
that considering heterogeneity in diffusion processes can have a big impact and to provide intuitions as to why.
Natural next steps would be to investigate the further implications of heterogeneity of activity in settings efficient vaccination policies, as well as enhancement of the diffusion of innovations by picking the best `seeds.'

Finally, here we examined the optimal structure of agents without constraints on the numbers of different types of bursty agents.  In some settings it may be possible to 
incentivize or choose agents to be bursty and others to be reversing.  A logical next problem to study is to consider a fixed, relatively small number of bursty agents (or intermediaries who are `always active') and pick the best `bursts' (or best `intermediaries') for the enhancement of the diffusion. This `optimal bursts' (or `optimal intermediaries') question also leads to several follow-ups: Is this a computationally hard problem? Would (as in \cite{kempe2003maximizing} for optimal seeds) a `greedy' algorithm perform well in approximating the optimal placement of bursty (or intermediary) agents? Similar to the exercise in \cite{banerjeecdj2013}, which centrality measures perform well in practice for finding the optimal placement of such agents? And, along the lines of \cite{akbarpour2017diffusion}, how many additional placements would have to be picked at random to prompt a larger diffusion than the optimum? 

\bibliographystyle{ecta}
\bibliography{networks_dyn2}

\section*{Appendix}\label{appendix}

\noindent {\bf Proof of Proposition \ref{three}:}
First, we show that it is never optimal to have the first agent be Sticky, either in terms of the expected number of infected agents or the probability of total infection.
This is clear in the case in which the second agent is Sticky, since then the probability that the second agent becomes infected is $\lambda^2$ if the
first agent is Sticky and is $\lambda( 1-(1-\lambda)^T)>\lambda^2$ if the first agent is Poisson.  Note that this implies that both the expected number of infected agents and the probability of total infection are higher by having the first agent be Poisson when the second agent is Sticky, since this is independent of what happens past the second agent conditional upon that agent being Sticky in both cases.
Next, consider the case in which the second agent is Poisson (and again, this applies for both both the expected number of infected agents and the probability of total infection).  Consider the $T$ periods in which the first agent might infect the second agent.
Let $X$ be the number of periods that the second agent is active out of those $T$.  If $X$ is 0 or 1, then having the first agent be Sticky or Poisson is equivalent.
However, if $X>1$, then the chance that the first agent is active in at least one of those $X$ periods is $\lambda$ for the Sticky agent and $1-(1-\lambda)^X>\lambda$ for the Poisson agent.
Thus, it is better to have the first agent be Poisson.

The following straightforward calculations
are useful in what follows.
The probability of a second node adjacent to a first one getting infected, conditional upon the first one being infected,
as a function of the configuration is:

\begin{eqnarray*}
PS&=& \lambda( 1-(1-\lambda)^T) \\
SP&=&  1-(1-\lambda)^T \\
PP&=&  1-(2\lambda(1-\lambda) +(1-\lambda)^2)^T  =  1- (1-\lambda^2)^T \\
SS&=& \lambda.
\end{eqnarray*}

To prove Proposition \ref{three}, first note that given the above expressions, $SP>PP>PS$
and
$SP>SS>PS$.

Let us next consider the configuration of three agents (with the first one randomly infected), and presume that the first agent is Poisson since we have already showed that to  be optimal.
Note that $PP>PS$ implies that $PPP>PPS$, and $SP>SS$ implies that  $PSP>PSS$, and moreover these comparisons hold
 {\sl both in terms of the last person being infected and the overall expected number of infections}.\footnote{Note that PPP and PPS lead to the same chances of the second member being infected and PPP has a conditional and unconditional higher expectation of the third member being infected, and so leads to a higher expectation.  The same is true of a comparison between PSP and PSS.}
Thus, the maximal string in terms of overall infection or expected number of infections is either PPP or PSP, as PSS and PPS are dominated (as is SSS by similar reasoning) - Let us compare those two.

First, let us do the comparison in terms of the probability of total infection.
For that $PPP= \left[ 1-(1-\lambda^2)^T\right]^2$ and
$PSP= \lambda \left[ 1- (1-\lambda)^T\right]^2$.

So,
$PSP> PPP$
if and only if
\begin{equation}
\label{leq}
\lambda> f(\lambda) = \left[\frac{ 1- (1-\lambda^2)^T}{ 1- (1-\lambda)^T} \right]^2.
\end{equation}
Next, we show that
$\lambda = \left[\frac{ 1- (1-\lambda^2)^T}{ 1- (1-\lambda)^T} \right]^2$ has a unique solution in $(0,1)$.
Consider the function $g(\lambda) = (1 - (1-\lambda)^T)^2$. It is easy to check that $g'(0) = g'(1) = 0$, $g$ is increasing in between, and $g''(0) > 0$ and $g''(1) = 0$ and that $g'''$ is negative. Therefore, $g(\lambda)$ is strictly convex at the beginning, and then becomes strictly concave, with a unique inflection point. Thus, $h(\lambda) = g(\lambda)/\lambda$ is monotonically increasing for $\lambda$'s below the inflection point and then monotonically decreasing after that. Also, $h(1) = 0$ and $\lim_{\lambda \to 0} h(\lambda) = 0$.

Next note that
solving for the fix point of $f(\lambda)$ is equivalent to
solving for $h(\lambda) = h(\lambda^2)$.
Since  $h(\lambda) $ is monotonically increasing for $\lambda$'s below the interior inflection point and then monotonically decreasing after that,with  $h(1) = 0$ and $\lim_{\lambda \to 0} h(\lambda) = 0$, it follows that
this equation has a unique solution in $(0, 1)$, denoted $\lambda^*$.

For small $\lambda$, $f(\lambda)$ is approximately $\left[\frac{T\lambda^2}{T\lambda}\right]^2 =\lambda^2$,
and so the (\ref{leq})  holds for small lambda, and so the condition holds for $\lambda<\lambda^*$ .

Next, to do the comparison in terms of infected agents, consider the expected number of infected agents beyond the first Poisson agent.
For that the expectations are:
\[
PPP= 2\left[ 1-(1-\lambda^2)^T\right]^2 + \left[ 1-(1-\lambda^2)^T\right](1-\lambda^2)^T  = \left[ 1-(1-\lambda^2)^T\right]\left[ 2-(1-\lambda^2)^T\right]
\]
and
\[
PSP= 2\lambda \left[ 1- (1-\lambda)^T\right]^2 + \lambda \left[ 1- (1-\lambda)^T\right](1-\lambda)^T = \lambda \left[ 1-(1-\lambda)^T\right]\left[ 2-(1-\lambda)^T\right].
\]
So,
$PSP> PPP$ in terms of the expected number of infections
if and only if
\[
\lambda>  \frac{ \left[ 1-(1-\lambda^2)^T\right]\left[ 2-(1-\lambda^2)^T\right]}{ \left[ 1-(1-\lambda)^T\right]\left[ 2-(1-\lambda)^T\right]} .
\]

The rest of the proof is similar to the previous analysis of the fixed point of an analogous function. \eproof

\bigskip

\noindent {\bf Proof of Proposition \ref{tree}:}

First,
the argument that the first agent should be Poisson is as in the previous proposition.  Similarly, the last agent being Poisson follows from the $PP>PS$ and $SP>SS$ from
the previous proposition.

Next, note that a comparison of the probability of total infection between $PSPSP\ldots SP$  or $PSSSS\ldots SP$ boils down to a comparison
of the probability of both $PS$ becoming infected vs $SS$, both following an infected $S$.
Those two calculations are
\[
PS: \ \lambda \left[ 1- (1-\lambda)^T\right]^2,
\]
and
\[
SS: \ \lambda^2.
\]
So, alternating is better if and only if
\[
\left[ 1- (1-\lambda)^T\right]^2> \lambda.
\]
Note that this is equivalent to $g(\lambda)>\lambda$, where $g(\lambda)$ was defined in the previous proof and  is initially convex and eventually concave and having a unique fixed point in $(0,1)$.   Note also that  $g'(0)=g'(1)=0$ and $g(0)=0$ and $g(1)=1$.  It follows that $g(\lambda)<\lambda$ for $\lambda<\lambda^{**}$ and then this reverses for $\lambda>\lambda^{**}$.
Thus, $g(\lambda)>\lambda$  if and only if $\lambda>\lambda^{**}$,
which establishes the comparison between $PSPSP\ldots SP$  or $PSSSS\ldots SP$ appearing in the proposition.

Next, note that when $g(\lambda)=\lambda$, then
$g(\lambda^2)<\lambda^2$, since $g(\lambda)<\lambda$ below $\lambda^{**}$.
This implies that $\lambda>
\left[\frac{ 1- (1-\lambda^2)^T}{ 1- (1-\lambda)^T} \right]^2$
from our proof of Proposition \ref{three}, which implies that
$\lambda<\lambda^*$, and so $\lambda^{**}<\lambda^*$.

Next, note that
a comparison between $PSPSP\ldots SP$  or $PPPPP\ldots PP$ boils down to a comparison between $SP$ and $PP$ following an infected $P$.  This is equivalent
to the calculation of $PSP$ versus $PPP$ from Proposition \ref{three}.

Let us then consider other possible sequences that involve beginning and ending $P$'s.

First, let us argue that it is not possible to have any instances of both $SS$ and $PP$ in the same sequence.
Consider a sequence that contains $SS$.
Since the sequence begins and ends with $P$'s, there must exist both $PSS$ and $SSP$ somewhere in the sequence.
This (generically) implies that $PSS>PPS$ and  $SSP>SPP$.  This means that there cannot exist a $PP$ in the sequence.
(If there were a $PP$, then since the sequence also has $S$s, somewhere there is at least one of $PPS$ or $SPP$.)

Next, let us argue that it is not possible to have $P$ in the interior of the sequence if there is some instance of $SS$.
First, given the odd number of interior nodes and the fact there there is no repetition of $P$'s in the sequence, if there is one instance of $SS$, there must be at least two such instances.    If there is a sequence of $SSS$, then it must (generically) dominate $SPS$, which contradicts the presence of an interior $P$.
If instead these two instances of $SS$ have a $P$ somewhere between them (possibly several, alternating with $S$s),
then there would be a sequence of the form $SSPSS$ or $SSPSPSS$, etc., somewhere.    Let us consider the first case, as the others are easy extensions.
The presence of $SSPSS$ means that $SPS$ dominates $SSS$, and so the value of $SP$ times $PS$ is larger than the value of $SS$ squared.
The value of $SSPSS$ is the values of $SS^2$ times $SP$ times $PS$.   Instead the value of $SPSPS$ is the values of $SP^2$ times $PS^2$, which is larger since the values of $SP$ times $PS$ is larger than the value of $SS$ squared.  Thus, we reach a contradiction.

A parallel argument implies that it is not possible to have any (interior) $S$s if there is some instance of $PP$.

Given the above arguments, the only remaining sequences have either all $S$s interior, all $P$'s interior, or fully alternate $S$ and $P$, which are the sequences we have already compared.

This completes the proof. \eproof

\bigskip

\noindent {\bf Proof of Proposition \ref{star}:}

The optimality of having Poisson nodes as an end node has already been established in Proposition \autoref{three}.
The probability of full contagion is the probability of the central node getting infected times the probability of the central node infects all other nodes. So the probability of full contagion when the central node is $P$ is:

$$ P_p = \Big(1 - (1-\lambda^2)^2)\Big) \Big(2\lambda(1-\lambda)\lambda^{n-1} + \lambda^2(1- (1- \lambda)^2)^{n-1}\Big) $$

And the probability of full contagion when the central node is $S$ is:

$$ P_s =  \Big(\lambda(1 - (1-\lambda)^2)\Big) \Big(1 - (1-\lambda)^2 \Big)^{n-1} = \lambda\Big(1 - (1-\lambda)^2 \Big)^{n}$$

Now we show that $lim_{n \to \infty}(P_p / P_s) < 1$, which then proves the proposition. Note that $(1- (1-\lambda)^2) = \lambda(2-\lambda)$. Thus,

$$lim_{n \to \infty}(P_p / P_s) = lim_{n \to \infty} \frac{\Big(1 - (1-\lambda^2)^2\Big)\Big(2\lambda^n(1-\lambda) + \lambda^{n+1}(2-\lambda)^{n-1}\Big)}{\lambda^{n+1}(2-\lambda)^n} $$

This can be simplified to:

$$lim_{n \to \infty}(P_p / P_s) = lim_{n \to \infty} \Big(1 - (1-\lambda^2)^2\Big)\Big(\frac{2(1-\lambda)}{\lambda(2-\lambda)^n} + 1/(2-\lambda) \Big) = \frac{(1 - (1-\lambda^2)^2)}{2-\lambda},$$

which is less than 1 for any $0 <\lambda < 1$. \eproof


\bigskip

\noindent {\bf Proof of \autoref{mustalternate}:}
Let $p^*=1-\lambda$ denote the Poisson $p$ associated with $\lambda$.

We provide the same proof for
the expected number of infected nodes or the overall probability of full contagion.
The proof is as follows.   Suppose, to the contrary, that the maximizer (either of the expected number of infections or the probability of overall infection) involves
all nodes having the same $p$.    We show that changing the $p$ for a leaf node will strictly increase the probability that the leaf node becomes infected (conditional on its neighbor being infected, as well as unconditionally).
Since changing the $p$ for a leaf node does not change the infection probability for any other node, this increases both the expected number of infections and the probability of overall infection.

We offer the proof for $T=2$, and examine the probability of a leaf node becoming infected conditional upon its predecessor being infected.

First, consider the extreme (Sticky) case of  $p=0$.  Note that $SP$ dominates $SS$ from our earlier analysis, and so in that case it is direct that
the leaf node should differ.

Next, consider a case in which $p\neq 0$.

First, note that $pp$ (recalling that $T=2$) is equal to
\[
(1-p)\lambda + pq \lambda + (1-p)^2 (1-\lambda) q,
\]
and noting that $q = \frac{p \lambda }{1-\lambda}$,
$pp$ can be written as
\begin{equation}
\label{pp}
\lambda \left[(1-p) + p^2\frac{\lambda}{1-\lambda}  + (1-p)^2 p\right].
\end{equation}
Thus, an optimal $p$ must maximize (\ref{pp}), and from the first order conditions
must satisfy
\[
0=-1 + 2p\frac{\lambda}{1-\lambda}  + 1-4p+3p^2 \Rightarrow \frac{3}{2}p^2 = p\left(2- \frac{\lambda}{1-\lambda}\right) ,
\]

which has two solutions:  $p=0$ and
\begin{equation}
\label{ppfoc}
p=  \frac{2}{3}\left(2- \frac{\lambda}{1-\lambda}\right).
\end{equation}

Taking the second derivative of (\ref{pp}), we find
\[
2\frac{\lambda}{1-\lambda}  -4+6p,
\]
which is only nonpositive if
\begin{equation}
p\leq \frac{1}{3}\left(2- \frac{\lambda}{1-\lambda}\right)
\end{equation}
and so the second solution from (\ref{ppfoc}) cannot be a maximizer.

Also, note that the other corner of $p=1$ cannot be a solution.  This follows directly since then (\ref{pp})
is equal to $\lambda^2/(1-\lambda) < \lambda$ (since $\lambda<1/2$), while (\ref{pp}) becomes $\lambda$ when $p=0$.

Thus, a solution of all the same $p$'s must have $p=0$, which we have already shown not to be possible.\eproof

\newpage

\section{Online Appendix}

\subsection{Additional Proofs}

\

\noindent {\bf Proof of Proposition \autoref{treeReverse}:}

The probability that a P is never active is $(1-\lambda)^T$.   The probability that an R is never active is
$(1-\lambda)(1-q)^{T-1}$ where $1-q<1-\lambda$, since a reversing node has a probability of staying in the inactive state that is lower than the overall probability of being in that state.   In particular,  $q = \min \{1, \lambda/ (1-\lambda)\}$, and so
 $1-q = \max \{ 0,  (1-2\lambda)/ (1-\lambda)\}$, and  $ (1-2\lambda)/ (1-\lambda) < 1-\lambda$ since
  $ 1-2\lambda < 1-2\lambda +\lambda^2$ given that $\lambda> 0$.

This implies that
$RS ...$  always beats $PS....$,   and
similarly   that
$.... SR$  beats $....SP$.

These facts imply that $RSPS...PSR$   and $RSSS...SSR$ dominate $PSPS...PSP$   and $PSSS...SSP$, respectively, and so this, together with Proposition \ref{tree}, implies that whenever $\lambda<\lambda^{*}$, the optimal configuration involves $R$.

Next, we focus on the case of $T=2$.   When $\lambda \leq 1/2$, an $R$ node has $p=1$ and $q=\lambda/(1-\lambda)$.

Let us first calculate the chance of the second node getting infected under various scenarios for the first two nodes (presuming that the first starts randomly):

$RR\ldots$:  $\lambda^2 + 2\lambda(1-\lambda) (1-p) q  + (1-\lambda)^2 q^2 =  2\lambda^2$

$RS\ldots$:  $\lambda^2 + \lambda(1-\lambda)  q  =  2\lambda^2$

$SR\ldots$:  $\lambda^2 + \lambda(1-\lambda)  q  =  2\lambda^2$

$RP\ldots$:  $\lambda^2 + \lambda(1-\lambda) (1-p) \lambda  + \lambda(1-\lambda)  q \lambda + (1-\lambda)^2 q \lambda =  2\lambda^2$

$PR\ldots$: $\lambda^2 + \lambda(1-\lambda) (1-p) \lambda  + \lambda(1-\lambda)  q \lambda + (1-\lambda)^2 q \lambda =  2\lambda^2$

$PS\ldots$:  $\lambda(\lambda + \lambda(1-\lambda)  ) = ( 2-\lambda)\lambda^2$

$SP\ldots$:  $\lambda(\lambda + \lambda(1-\lambda)  ) = ( 2-\lambda)\lambda^2$

$SS\ldots$:  $\lambda^2$

$PP\ldots$:  $\lambda^2 + \lambda^2(1-\lambda^2)  ) = ( 2-\lambda^2)\lambda^2$

From these calculations, and noting that it is strictly best to have the second node be $S$ (since then it is active for sure), the unique optimal starting configuration is $RS\ldots$.

Note now, that this implies that the second node is active conditional on having been infected since it is $S$.

Next, let us consider nodes beyond the first two.

We analyze the probability that a subsequent node gets infected conditional upon a previous one being infected (once we are past the initial node).
Here the calculations lead to:

 $\ldots PR\ldots$: $\lambda^2 + \lambda(1-\lambda) (1-p) \lambda  + \lambda(1-\lambda)  q \lambda + (1-\lambda)^2 q \lambda =
  \lambda^3 + (2-\lambda) \lambda^2=2\lambda^2$

 $\ldots PP\ldots$: $\lambda^2 + \lambda^2(1-\lambda^2)  = ( 2-\lambda^2)\lambda^2$

 $\ldots PS\ldots$:  $\lambda(\lambda + \lambda(1-\lambda)  ) = ( 2-\lambda)\lambda^2$

 $\ldots SR\ldots$:  $(1-p)  + p q = \lambda+(1-\lambda)q= 2\lambda  $

 $\ldots SP\ldots$:  $\lambda + \lambda(1-\lambda)  = ( 2-\lambda)\lambda$

 $\ldots SS\ldots$: $\lambda$

$\ldots RR\ldots$: $(1-p) [ \lambda +  (1-p)(1- \lambda)q]   + p q \lambda  = \lambda^2/(1-\lambda) $

$\ldots RS\ldots$: $(1-p) \lambda  + p q \lambda  = \lambda^2/(1-\lambda)$

$\ldots RP\ldots$: $(1-p)\lambda[ 1 +  (1-p)(1- \lambda)]  + p q \lambda  =  \lambda^2/(1-\lambda)$

Note that these imply that the best last node is $R$ (with a note that if the second to last node is also $R$, then we are indifferent as to the last node).
For the arguments below, we thus take the last node to be $R$ - and show the unique optimal configuration given that, and then we can vary the last node if the second to last node happens to be $R$, which it never does.

Using the above calculations, let us next show that it is never optimal to have an interior $P$.

First, from the above calculations and $\lambda<1/2$, it is easy to check that conditional on the first node being infected $..PSS>.. PPS$, $..PSR>..PPR$, and $..PSP>..PPP$.
(To see the last one, note that  $(2-\lambda)^2 \lambda^3> (2-\lambda^2)^2\lambda^4$,  since it is equivalent to
$(2-\lambda)^2> (2-\lambda^2)^2\lambda$, and noting that the right hand side is smaller than $2-\lambda^2$, which is less than the left hand side which is
$4-4\lambda+\lambda^2$.)
These imply that there will never be two $P$'s in a row.
The only other ways in which $P$ could enter in the interior (without having two $P$'s in a row) is either as $..SPS$, $..SPR$, $..RPS$, and $..RPR$.
Note that these are less than $..SRS$, $..SSR$, $..RSS$, and $..RSR$, respectively.

Thus, the optimal configuration involves only $R$ and $S$ (except if the second to last node is $R$ in an optimal configuration, in which case any last node is optimal, but that case will not arise).

Before examining the optimal intermediate patterns, let us consider the case in which $\lambda>1/2$.

\smallskip

Next, we focus on the case of $T=2$ and $\lambda>1/2$, so that for the $R$ nodes: $p=(1-\lambda)/\lambda$ and $q=1$ and $1-p= (2\lambda-1)/\lambda$.

Let us first calculation what the chance of the second node getting infected under various scenarios for the first two nodes (presuming that the first starts randomly):

$RR\ldots$:  $\lambda^2 + 2\lambda(1-\lambda) (1-p) q  + (1-\lambda)^2 q^2 = \lambda + (1-\lambda)( 2\lambda-1) $

$RS\ldots$:  $\lambda^2 + \lambda(1-\lambda)  q  =  \lambda$

$SR\ldots$:  $\lambda^2 + \lambda(1-\lambda)  q  =  \lambda$

$RP\ldots$:  $\lambda^2 + \lambda(1-\lambda) (1-p) \lambda  + \lambda(1-\lambda)  q \lambda + (1-\lambda)^2 q \lambda = \lambda^2 (3-2\lambda) $

$PR\ldots$: $\lambda^2 + \lambda(1-\lambda) (1-p) \lambda  + \lambda(1-\lambda)  q \lambda + (1-\lambda)^2 q \lambda = \lambda^2 (3-2\lambda) $

$PS\ldots$:  $\lambda(\lambda + \lambda(1-\lambda)  ) = ( 2-\lambda)\lambda^2$

$SP\ldots$:  $\lambda(\lambda + \lambda(1-\lambda)  ) = ( 2-\lambda)\lambda^2$

$SS\ldots$:  $\lambda^2$

$PP\ldots$:  $\lambda^2 + \lambda^2(1-\lambda^2)  ) = ( 2-\lambda^2)\lambda^2$

From these calculations,
we can conclude that $PS\ldots$,  $SP\ldots$, $SS\ldots$,  and $SR\ldots$,  are all dominated by $RS\ldots$ (noting that it is always best to have a second node be $S$ conditional upon it being infected since it will then stay active).

So the possible starting cases are $RR\ldots$, $RS\ldots$, $PR\ldots$, $PP\ldots$, and $RP\ldots$.
Next, note that from the expressions above $RR\ldots$ dominates $PR\ldots$ \footnote{Here note that the difference is $2\lambda^3-5\lambda^2+4\lambda-1 = (2\lambda-1)(1-\lambda)^2>0$.}
and $RP\ldots$ dominates $PP\ldots$, and so we are down to
$RR\ldots$, $RS\ldots$, and $RP\ldots$ as starting.\footnote{Although these may also be ordered, how the second node interacts with subsequent nodes differs across these three
starting configurations.
So, we cannot rule any of them out until we do further calculations about the interaction with subsequent nodes.}

We can then analyze the probability that a subsequent node gets infected conditional upon a previous one being infected (once we are past the initial node).
Here the calculations lead to:

 $\ldots PR\ldots$: $\lambda^2 + \lambda(1-\lambda) (1-p) \lambda  + \lambda(1-\lambda)  q \lambda + (1-\lambda)^2 q \lambda =(3-2\lambda)\lambda^2$

 $\ldots PP\ldots$: $\lambda^2 + \lambda^2(1-\lambda^2)  = ( 2-\lambda^2)\lambda^2$

 $\ldots PS\ldots$:  $\lambda(\lambda + \lambda(1-\lambda)  ) = ( 2-\lambda)\lambda^2$

 $\ldots SR\ldots$:  $(1-p)  + p q =  1 $

 $\ldots SP\ldots$:  $\lambda + \lambda(1-\lambda)  = ( 2-\lambda)\lambda$

 $\ldots SS\ldots$: $\lambda$

$\ldots RR\ldots$: $(1-p) [ \lambda +  (1-p)(1- \lambda)q]   + p q \lambda  = \lambda +(1-\lambda)(2\lambda -1)^2/\lambda^2$

$\ldots RS\ldots$: $(1-p) \lambda  + p q \lambda  = \lambda$

$\ldots RP\ldots$: $(1-p)\lambda[ 1 +  (1-p)(1- \lambda)]  + p q \lambda  = \lambda + (1-\lambda)(2\lambda -1)^2/\lambda $

These all make it clear that the last node should be an $R$ as well.  So, we only need to investigate the intermediate patterns.

We first argue that in any entry (except possibly the last entry), regardless of what comes before or after, it is better to have $R$ or $S$ compared to $P$.

First, we show that $R$ is always a better receiver than $P$.

First, note that in terms of being the second node (noting that $R$ is always the starting node), $RR\ldots$ has a higher probability that $RP\ldots$.   To see this, note that the difference in probabilities can be written
as $2\lambda^3 + 4\lambda-5\lambda^2 -1 =  (2\lambda-1)(1-\lambda)^2 >0$.

Similar sorts of comparisons show that
$\ldots SR\ldots$ has a higher probability than $\ldots SP\ldots$,
$\ldots RR\ldots$ has a higher probability than $\ldots RP\ldots$, and
$\ldots PR\ldots$ has a higher probability than $\ldots PP\ldots$.

Next, note that $S$ is always a better sender node:
$\ldots SS\ldots$ has a higher probability than $\ldots PS\ldots$,
$\ldots SR\ldots$ has a higher probability than $\ldots PR\ldots$, and
$\ldots SP\ldots$ has a higher probability than $\ldots PP\ldots$.

$\ldots RS\ldots$ has a higher probability than $\ldots PS\ldots$

So since $R$ is a better receiver, there is no configuration with a $PS$ in it.

We are left with configurations of the form $PP$  or $PR$ appearing somewhere.

$PR$ could potentially come in 3 forms (given that we have to consider the $P$ as sandwiched):
$\ldots SPR\ldots$, $\ldots PPR\ldots$, $\ldots RPR\ldots$.

It follows from our calculations that
$\ldots SPR\ldots$ and $\ldots PPR\ldots$ are dominated by $\ldots SRR\ldots$ and $\ldots PRR\ldots$, respectively.
Also $\ldots RPR\ldots$ is dominated by $\ldots RSR\ldots$ for $\lambda\leq \lambda''' $  and by $\ldots RRR\ldots$ for $\lambda \geq  \lambda'''$,
where $\lambda'''\in (1/2,1)$ solves $\ldots RSR\ldots =\ldots RRR\ldots $.\footnote{The calculations to verify some of these various expressions are quite involved, but can be accomplished in some cases by comparing the expressions at various extreme values of $\lambda$ and then checking that the functions stay ordered by checking that intersections occur outside of the relevant interval of $\lambda$s, or that the derivative of the difference does not change signs.  It easiest to simply graph the functions in a program and check that they are properly ordered.}

$PP$ could conceivably come embedded in 3 triples  (noting that we already ruled out other combinations, for instance we showed there is no trailing $S$ and $\ldots PPR\ldots $ is has also been handled above):
$\ldots SPP\ldots$, $\ldots RPP\ldots$, $\ldots PPP\ldots$.

It then follows from comparing the expressions that
$\ldots SPP\ldots$, $\ldots RPP\ldots$ and  $\ldots PPP\ldots$ are dominated by
$\ldots SRP\ldots$, $\ldots RRP\ldots$ and $\ldots PRP\ldots$, respectively.

This implies that the only entries in an optimal configuration (either in terms of maximizing the expected number of infections or the probability of total diffusion) are $S$ and $R$.

Given that we only need to consider $S$ and $R$ entries, the remainder of the proof
parallels that of the proof of Proposition \ref{tree}, showing that we end up with either
$RSSS...R$ or $RSRS...R$.    Note that the second to last node is always an $S$, and so the unique last node is $R$.

There is one small change from that proof.   The starting node is presumed to be independently active in the first period, while an interior $R$ node once it is infected in one period is likely to be inactive.   That means that the calculations differ between a first node and an interior node when they are $R$ nodes.
The parallel to the proof of Proposition \ref{tree} can be used to show that past the third node it is best to have either $SSSS...R$ or $RSRS...R$.
The configuration for the first three nodes may be all $R$'s.
\eproof

\subsection{Details regarding the simulations pictured in Figure \ref{fig:heterog}.}\label{sim.details}

The simulation of Figure \ref{fig:heterog} is for an Erd\H{o}s-Renyi random graph with 20 nodes, with $\lambda=0.2$ and $p=0.25$. Sticky nodes are chosen randomly. One random node is infected and we count the number of realizations (out of 200,000) in which full diffusion happens.

\subsection{Additional figures illustrating results and simulations}

\begin{figure}[h!]
\centering
\subfloat[]{
\label{fig:Chains-1}
\includegraphics[width=0.45\textwidth]{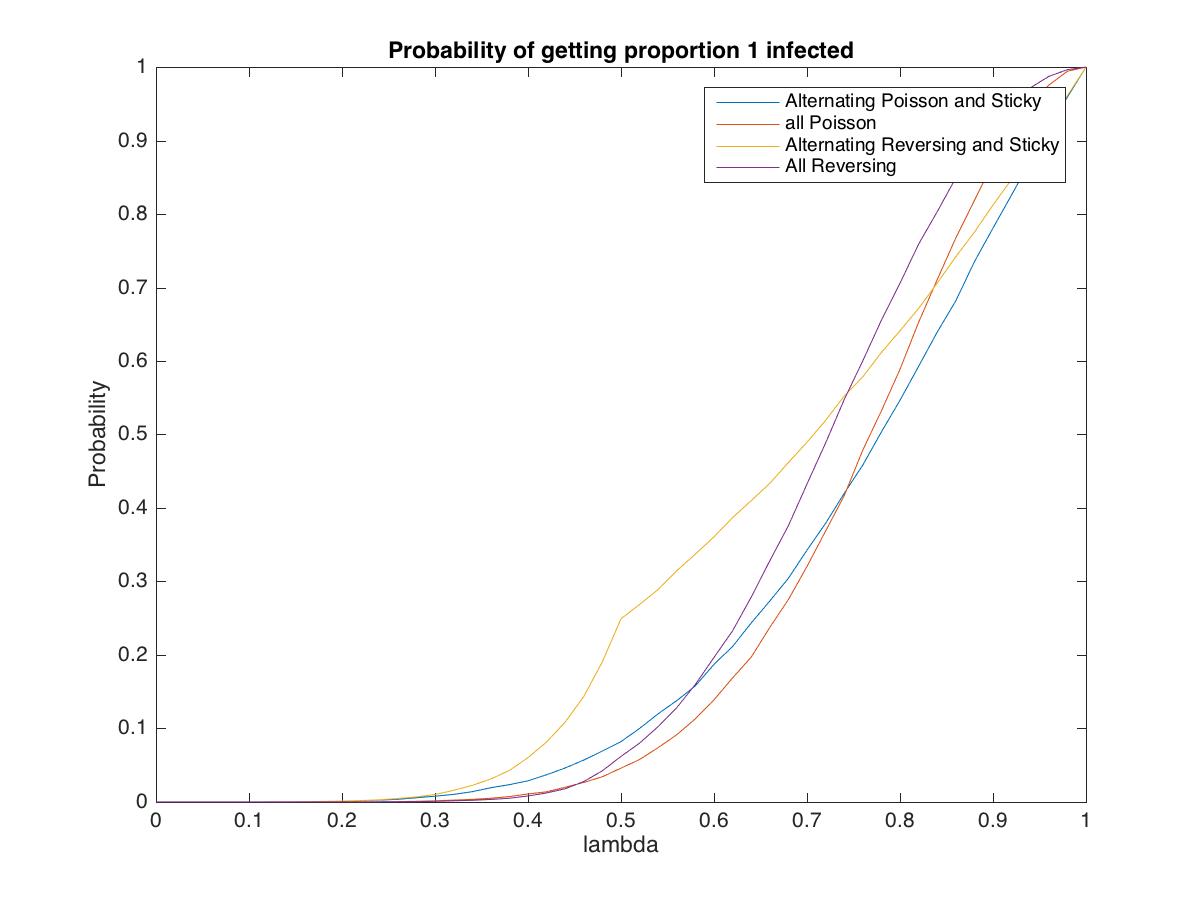}
}
\subfloat[]{
\label{fig:ChainsFrac}
\includegraphics[width=0.45\textwidth]{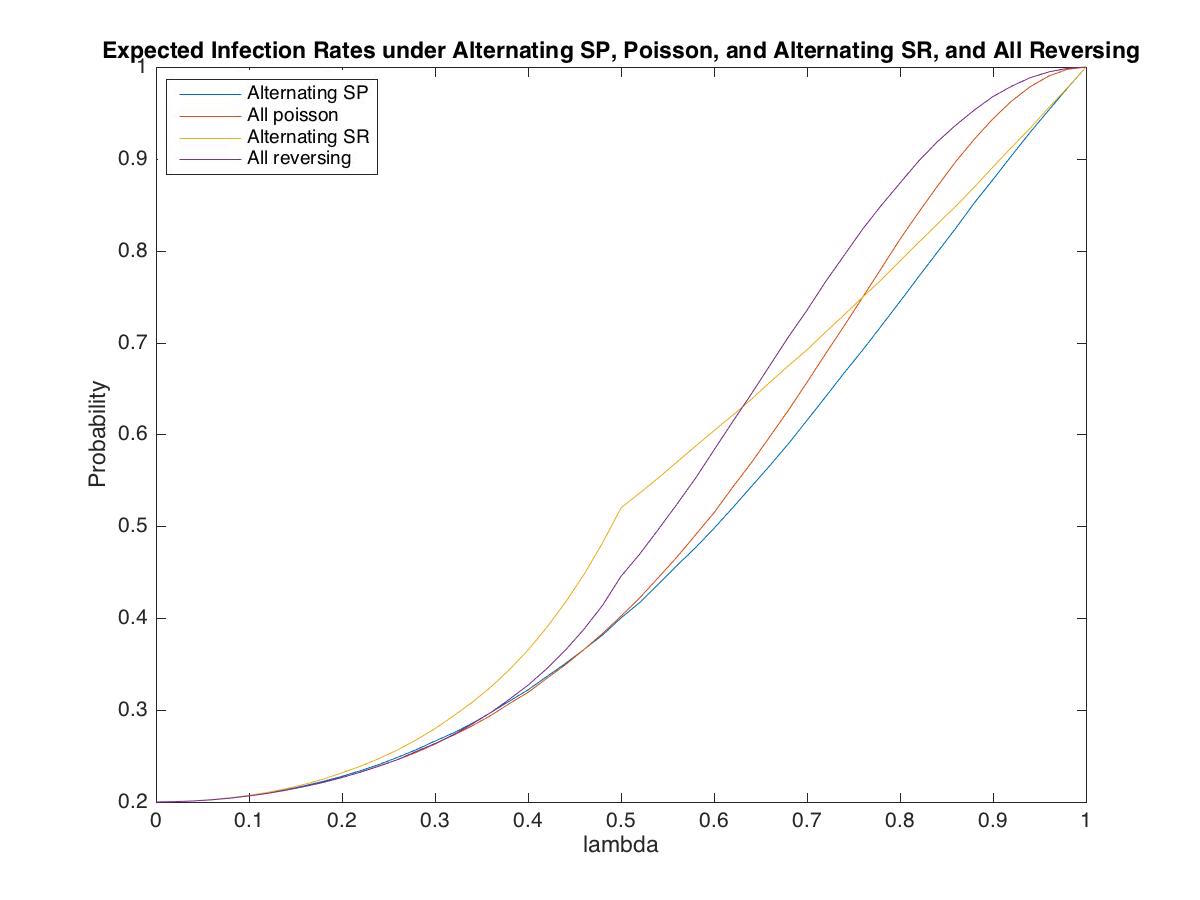}
}
\caption{\label{figChains2}Comparisons of infection probabilities and the average fraction infected under various configurations of nodes  on lines of five nodes with one randomly infected: (a) the probability of Getting All Nodes Infected, (b) the average Fraction of Nodes Infected.
}
\end{figure}

\begin{figure}[h!]
\centering
\subfloat[]{
\label{fig:ChainsRatios}
\includegraphics[width=0.45\textwidth]{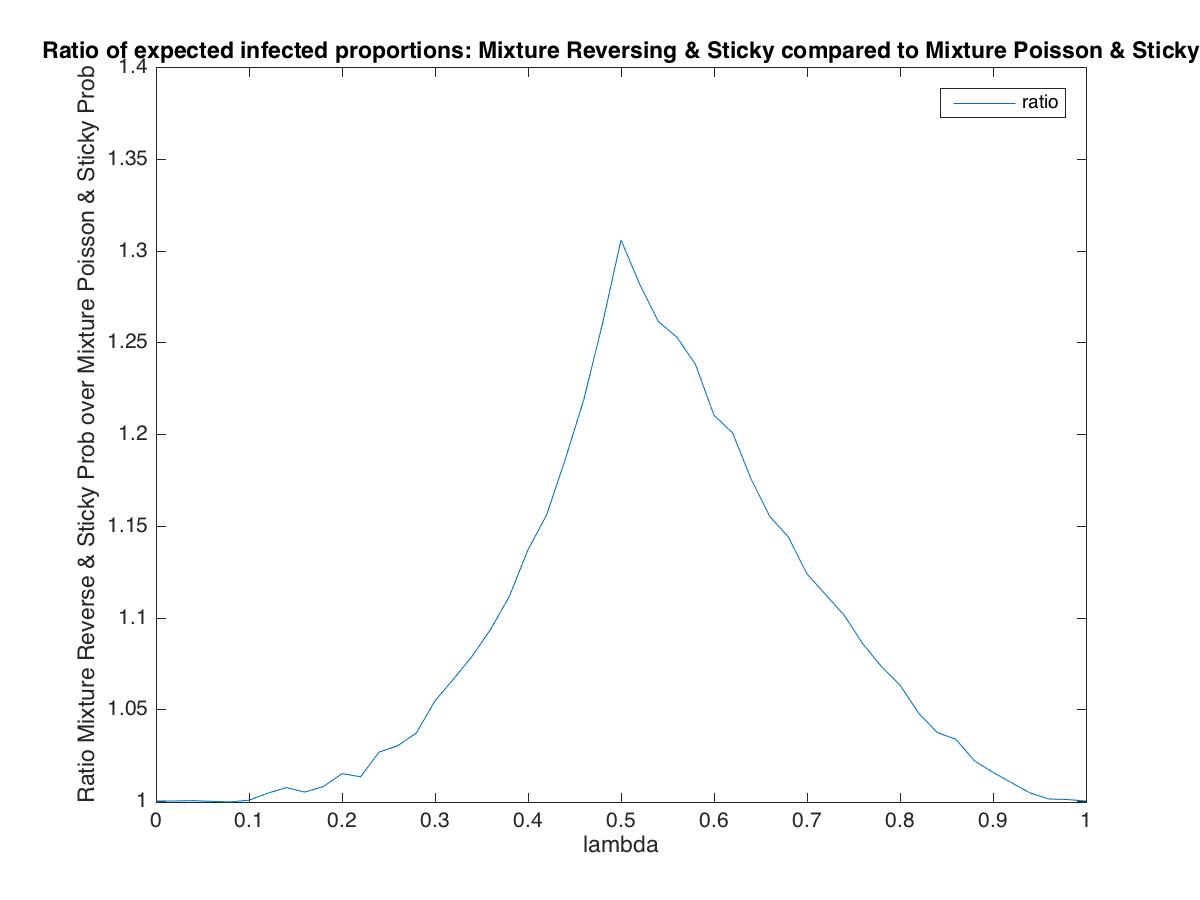}
}
\subfloat[]{
\label{fig:Chains5}
\includegraphics[width=0.45\textwidth]{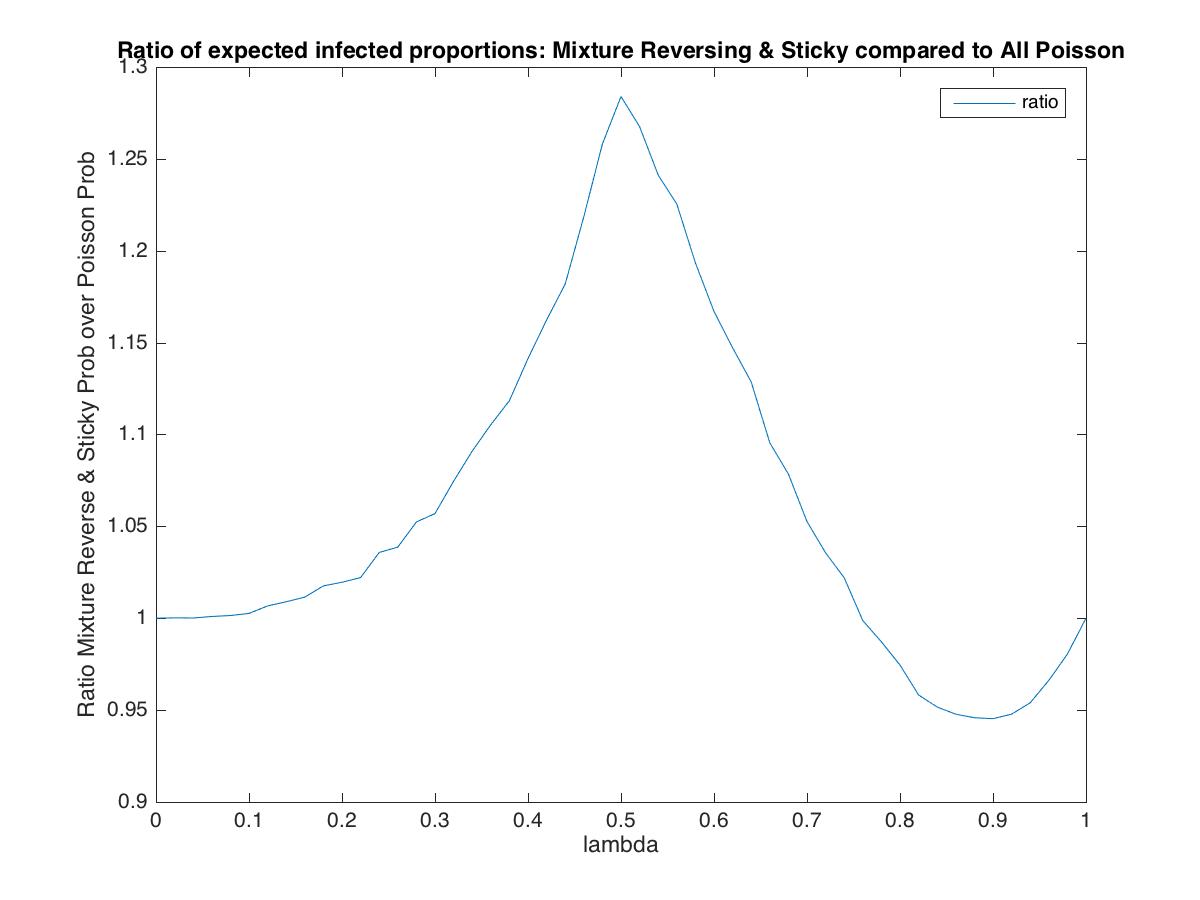}
}
\caption{\label{figChains1}
The ratio of expected infected proportions under various configurations of nodes on lines of five nodes with one randomly infected:
(a) alternating Reversing and Sticky compared to (over) alternating Poisson and Sticky, (b) alternating Reversing and Sticky compared to (over) all Poisson.}
\end{figure}

\begin{figure}[h!]
\centering
\subfloat[]{
\label{fig:ER-Reverse-1}
\includegraphics[width=0.45\textwidth]{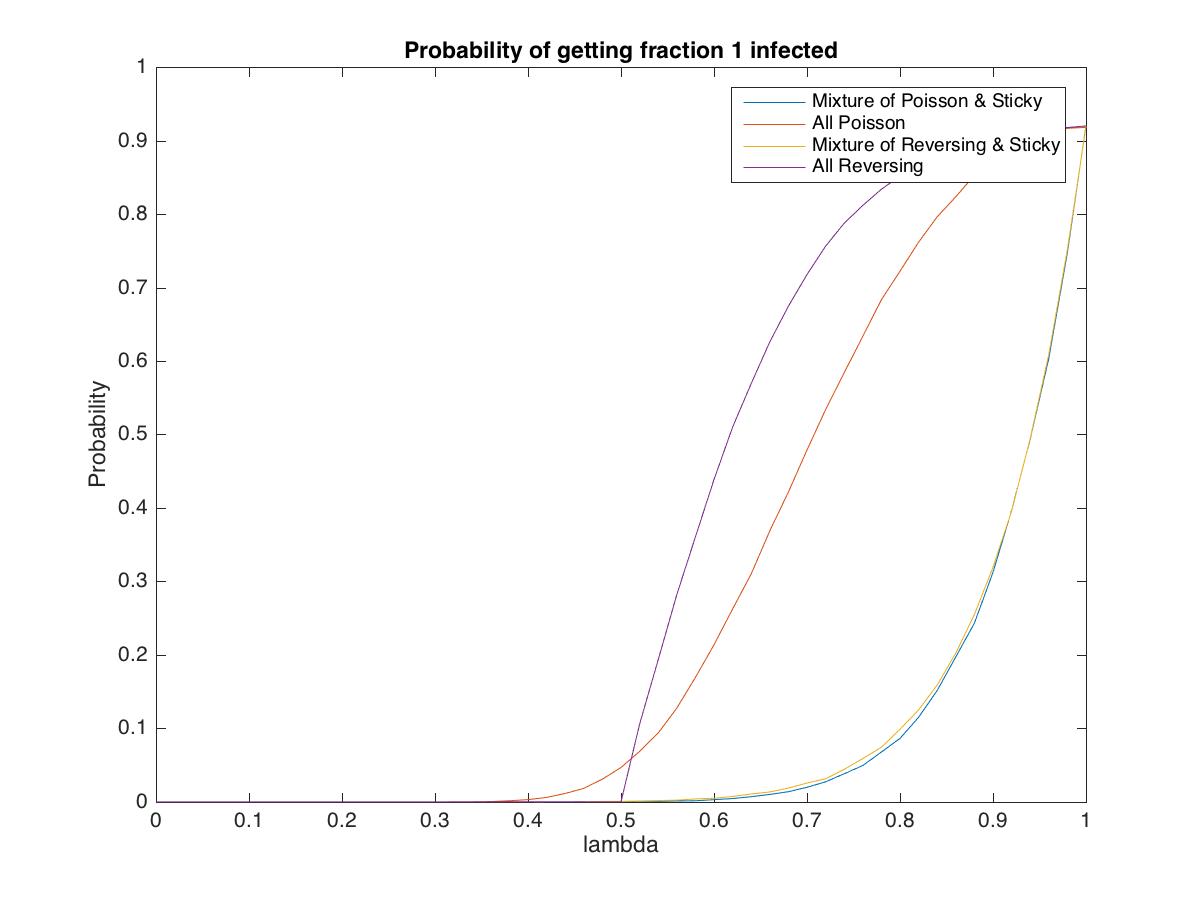}
}
\subfloat[]{
\label{fig:ER-ReverseAvg}
\includegraphics[width=0.45\textwidth]{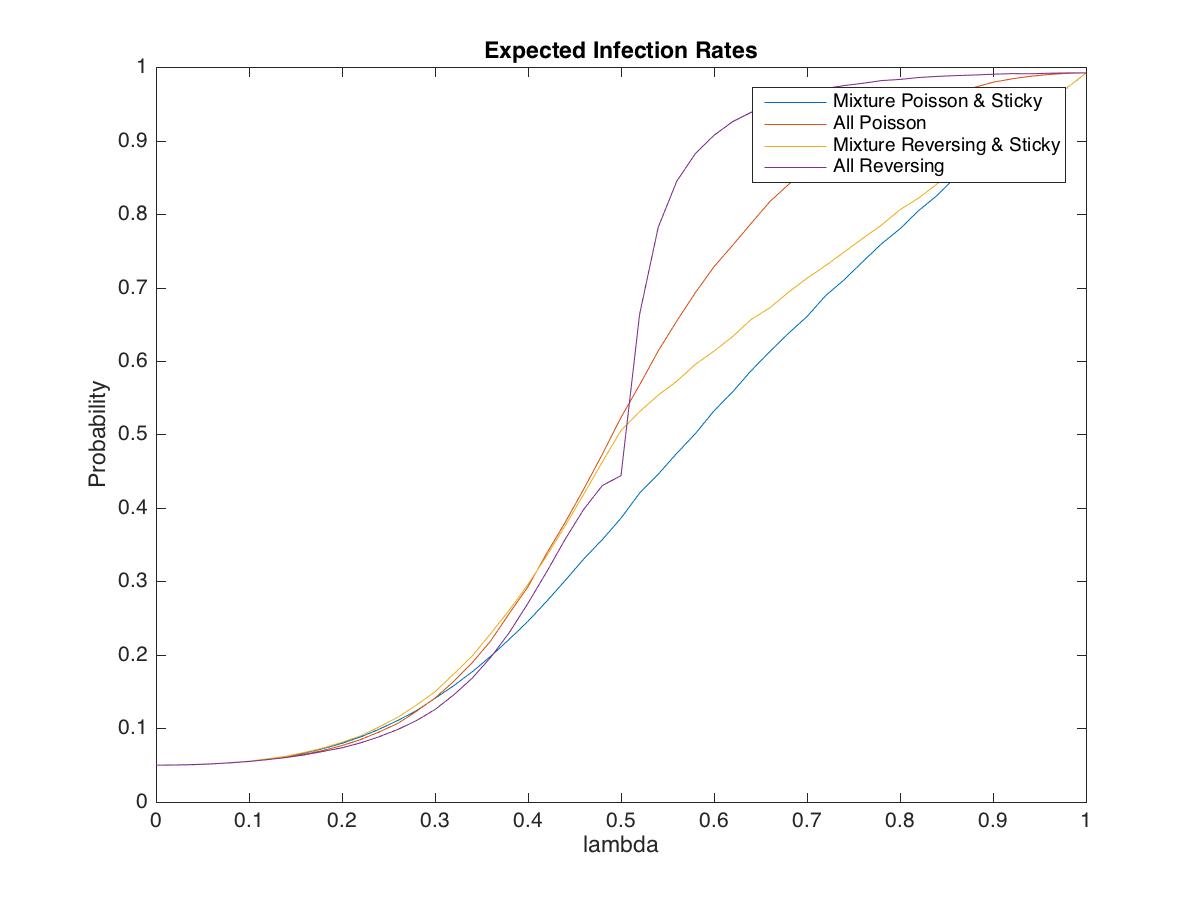}
}
\caption{\label{fig:ER-Reverse2}  Comparing all Poisson to an alternation of Poisson with Sticky to an alternation of Reversing nodes with Sticky nodes.
Erd\H{o}s Renyi Random networks on 20 nodes: (a) the probability of Getting All Nodes Infected, (b) the average Fraction of Nodes Infected.
}
\end{figure}

\end{document}